\documentclass[prd,11pt,reprint]{revtex4-1}

\usepackage{amsmath}    
\usepackage{graphicx}   
\usepackage{verbatim}   
\usepackage{color}      
\usepackage{hyperref}   
\usepackage{mathrsfs}
\usepackage{upgreek}
\usepackage{braket}
\usepackage{physics}
\usepackage{amssymb}

\usepackage{subfig}
\usepackage{float}
\usepackage{booktabs}

\newcommand{\ham}[0]{\mathcal{H}}

\begin{document}

\title{Semiclassical and Quantum Polymer Effects in the Flat Isotropic Universe}
\author{G. Barca, P. Di Antonio, G. Montani, A. Patti}
\affiliation{ENEA, Fusion and Nuclear Safety Department, C.R. Frascati, Via E. Fermi 45 (00044) Frascati (RM), Italy \\
Department of Physics, “Sapienza” University of Rome, P.le Aldo Moro, 5 (00185) Roma, Italy}

\date{February 2019}

\begin{abstract}
We analyze some relevant semiclassical and quantum features of the implementation of Polymer Quantum Mechanics to the phenomenology of the flat isotropic Universe.

We firstly investigate a parallelism between the semiclassical polymer dynamics of the flat isotropic Universe, as reduced to the effect of a modified simplectic structure, and the so-called Generalized Uncertainty Principle. We show how the difference in the sign of the fundamental Poisson bracket is reflected in a sign of the modified source term in the Friedmann equation, responsible for the removal of the initial singularity in the polymer case and for the survival of a singular point in the Universe past, when the Generalized Uncertainty Principle is concerned. 

Then, we study the regularization of the vacuum energy of a free massless scalar field, by implementing a second quantization formalism in the context of Polymer Quantum Mechanics. We show that, from this reformulation, naturally emerges a Cosmological Constant term for the isotropic Universe, whose value depends directly on the polymer parameter of the regularization. 

Finally, we investigate the behaviour of gravitational waves on the background of a modified dynamics, according to the semiclassical Friedmann equation. We demonstrate that the presence of a Bounce in the Universe past, naturally removes the divergence of the gravitational wave amplitude and they can, in principle, propagate across the minimum volume turning point. This result offers the intriguing perspective for the detection of gravitational signals coming from the pre-Big Bounce collapsing Universe. 
\end{abstract}

\maketitle

\section{Introduction} \label{introduction}
Modern Quantum Gravity approaches are mainly based on the use of Ashtekar-Barbero-Immirzi variables \cite{art:ash,art:rovelli}, which constitute the starting point for the construction of Loop Quantum Gravity theory \cite{art:immirzi,book:canonicalquantumgravity}. The main success obtained by this reformulation of the quantum gravitational field morphology is, on a phenomenological point of view, the derivation of a Big Bounce Cosmology \cite{art:ashbounce1,art:lqg3}. In fact, despite the minisuperspace model associated to homogeneous cosmological Universes \cite{book:primordialcosmology} prevents a full implementation of the $SU(2)$ symmetry, at the ground of the discretization of the geometrical operators (areas and volumes) \cite{art:cianfrani1,art:cianfrani2}, a notion of cut-off on the Universe volume and then a maximum critical density for the Planck era is recovered with a suitable procedure, recovering the general theory prescription and formalisms. 

Actually, the regularization procedure of the minisuperspace dynamics allows the construction of semiclassical equations for the Universe evolution, which turns out to be closely related to the metric approach in the polymer representation of canonical quantization \cite{art:corichi}. In  particular, in \cite{art:claudia}, the cubed scale factor, i.e. the Universe volume, has been identified as the natural variable in which the correspondence between the two semiclassical theories (i.e. semiclassical loop Cosmology and semi-classical polymer dynamics) are better linked to each other. In fact, for such a choice, the polymer parameter (the discretization step of the cubed scale factor) turns out to be directly linked to the Immirzi parameter. The peculiarity of such a configurational variable choice relies on the possibility to define a critical density depending on fundamental constants only, exactly like in Loop Quantum Cosmology. For a discussion on the use of the cubed scale factor in more general cosmological models, like the generic inhomogeneous solution, see \cite{art:antonini} and on the different phenomenological issue in the Mixmaster chaos of Bianchi IX, see \cite{art:crino}. 

In this paper, using the Polymer Quantum Mechanics framework, we study the phenomenology of the flat Robertson-Walker geometry, by adopting the cubed scale factor as  configurational variable.

In particular, we consider three different questions: one of conceptual relevance about the meaning of Polymer Quantum Mechanics on a perturbative level, and two phenomenological implications concerning the vacuum energy of a massless scalar field (i.e. the value of the Cosmological Constant as vacuum energy) and the propagation of gravitational waves through the Big Bounce. 

Firstly, we study the semiclassical polymer dynamics on a perturbative limit, when the cut-off parameter is small enough and the modified Hamiltonian formulation can be restated as a standard Hamiltonian constraint associated to modified Poisson brackets. This analysis puts the semiclassical polymer dynamics on the same level of the so-called Generalized Uncertainty Principle \cite{art:kempf}, with the non-trivial difference of a sign in the right-hand side of the brackets \cite{art:taub}. The polymer and Generalized Uncertainty Principle reformulations can then be seen as the phenomenological low energy limit of, respectively, Loop Quantum Cosmology \cite{art:ashcrino} and String Theory \cite{art:kempf,art:brane1,art:brane2,art:brane3}, implemented as a modified simplectic structure.

We study the behaviour of the flat isotropic Friedmann-Robertson-Walker Universe (dominated by radiation and stiff matter respectively) for both the two cases mentioned above, demonstrating that, while in the polymer approach the singularity is still removed as in the exact (non-perturbative) case, the Generalized Uncertainty Principle dynamics is still associated with a Big-Bang, i.e. the cosmological singularity survives. Furthermore, analyzing the structure of the Friedmann equation in these same cases, we show that they coincide respectively with those of exact polymer semiclassical mechanics (i.e. semiclassical Loop Quantum Cosmology) and Brane Cosmology approach, when the Universe density is sufficiently small with respect to the critical one.

Hence, we restrict our attention to the exact polymer quantum physics only and, as a first step, we study the vacuum energy of a massless scalar field living on the flat isotropic Universe. Despite a complete scheme of second quantization of the field is forbidden because it is no longer possible to define suitable creation and annihilation operators, we demonstrate that, due to polymer regularization, the vacuum energy no longer diverges. This result provides a non-zero Cosmological Constant to the Universe dynamics, whose value is however dependent on the value of the discretization parameter. A discussion of the possible fine-tuning required to deal with a 'dark energy candidate' is developed, but, for a Planckian discretization step, the resulting values of three dimensionless parameter of the model are still very peculiar. This investigation on the vacuum state of a free massless scalar field has an important conceptual value, since it demonstrates that a Cosmological Constant rigorously emerges in the Polymer Quantum Mechanics framework. However, we do not identify a mechanism for the reduction of the Cosmological Constant value to the actual one. A qualitative implementation of the upper limits for a Polymer cut-off on the physical space \cite{art:polbound} provides a vacuum energy density many orders of magnitude greater than the one requested by the Universe acceleration. We can only stress that the calculated ground state of the scalar field Hamiltonian function is not a state for the quantum dynamics of the system and, therefore, we dynamically have to deal with a time dependent expectation value on the vacuum state of the scalar field. 

The analysis of the gravitational wave propagation on a Bounce Cosmology offers a new point of view on the possibility to observe pre-Big Bounce features. In fact, the presence of the Bounce regularizes the wave amplitude which no longer diverges as in the Big-Bang model. Thus, it is, in principle, possible that gravitational waves produced in the collapsing Universe remain in linear regime across the Bounce and they could be today detected. In particular, we study the polymer deformation of the wave spectrum and consider the propagation of peaked wave packets, i.e. burst signals. It is interesting to notice that, in the limit of wavelengths of the ripples that are large in respect to the cut-off parameter, the standard properties \cite{book:weinberg} are recovered. In other words, the gravitational wave morphology is sensitive to the discretization parameter, although it lives in the configurational space and not in the physical one. The point is that the discretization of the scale factor of the Universe is clearly reflected on the nature of any physical spatial scale, including the physical wavelength. 

The paper is structured as follows. In Section \ref{standard cosmology} we present the standard FLRW cosmological model in its hamiltonian formulation, using as configurational variable the volume $V=a^3$. In Section \ref{polymer} we introduce the polymer representation of Quantum Mechanics in the momentum polarization; moreover we present a way to treat both quantum and semiclassical states in the polymer framework. In Section \ref{Gabriele} we analyze the FLRW Universe in the polymer framework, firstly with an 'exact' approach, and secondly in a perturbative approach, confronting the latter with the Generalized Uncertainty Principle approach. In Section \ref{Paolo} we recall the theory of the quantum harmonic oscillator in polymer representation and then we evaluate the vacuum energy density for a massless scalar free field in a flat FLRW background in the polymer framework. In Section \ref{Alberto} we discuss how, in terms of the semiclassical formulation of Polymer Quantum Mechanics, the introduction of a cut-off regularizes the amplitude of gravitational waves propagating through a flat FLRW Universe. We also study the spectrum of such waves and the time evolution of a Gaussian wave packet and we compare it to the classical case. Finally, in Section \ref{conclusions} we sum up the main result of the paper with concluding remarks.

\section{Standard Cosmology} \label{standard cosmology}
The Standard Cosmological Model (SCM) relies on the Friedmann-Robertson-Walker (FLRW) isotropic and homogeneous expanding Universe. This model is based on the Cosmological Principle (i.e. at large scales the Universe is isotropic and homogeneous, as confirmed by the CMB spectrum), on the perfect fluid approximation of the matter-energy content and on General Relativity, and its geometry is described by the Robertson-Walker metric:
\begin{equation}
ds^2=dt^2-a^2(t)\left[\frac{dr^2}{1-Kr^2}+r^2 d\theta^2+r^2 \sin^2{\theta} d\varphi^2\right]
\label{RWmetric}
\end{equation}
where $a(t)$ is the scale factor through which the whole expansion history of the Universe is parametrized. Our study will be focused on the flat Universe, so from here on we will use $K=0$. Besides, we will use natural units $\hslash=c=1$.

\subsection{Cosmological dynamics}
The evolution of the Universe is described by cosmological equations, which are derived by using the metric \eqref{RWmetric} in the Einstein equations, together with the energy-momentum tensor of the perfect fluid $T_{\mu\nu}^\text{PF}=\text{diag}(\rho,-P,-P,-P)$.\newline
The $00-$component of Einstein equations results in the first Friedmann equation:
\begin{gather}
H^2=\frac{\dot{a}^2}{a^2}=\frac{\chi}{3}\rho
\label{F1std(a)}
\end{gather}
that describes the relative velocity of the expansion as function of the matter-energy density of the Universe. Here $\chi=8\pi G$ is the Einstein constant.\newline
The $jj-$components are all equivalent and, combined with eq. \eqref{F1std(a)}, reduce to the second Friedmann equation, also known as acceleration equation:
\begin{gather}
\label{F2std(a)}
2\frac{\ddot{a}}{a}=-\frac{\chi}{3}(\rho+3P)
\end{gather}
that describes the relative acceleration of expansion.\newline
Finally, combining these two Friedmann equations, the continuity equation is obtained:
\begin{gather}
\dot{\rho}=-3\frac{\dot{a}}{a}(\rho+P)
\label{continuity(a)}
\end{gather}
It can be solved by using a polytropic constant equation of state $P=\omega\rho$, where $\omega$ is a parameter that can take values in the interval $\omega\leq1$ (greater values would result in a superluminal sound velocity and are therefore non physical). The solution is $\rho(a)=\rho_0\,a^{-3(1+\omega)}$ and it holds even with the semiclassical modifications that we will apply in the following chapters \cite{art:claudia}.

These three equations completely describe the dynamics of the Universe. Actually, since any one of the three can be derived from the other two, only two of them are strictly necessary. Usually in literature the first Friedmann and the continuity equations are chosen.\newline

\paragraph*{The content of the Universe} The matter-energy density $\rho$ that appears in the first Friedmann equation \eqref{F1std(a)} receives contributions from different kinds of cosmological fluids, each characterized by its own value of the polytropic parameter $\omega$:
\begin{equation}
\rho(a)=\sum_i\rho_i(a)=\sum_i \rho_0\,a^{-3(1+\omega_i)}
\end{equation}
where the subscript $i$ indicates the type of fluid: we have $\omega_\text{sm}=1$ for \emph{stiff matter}, $\omega_\text{r}=\frac{1}{3}$ for radiation, $\omega_\text{m}=0$ for baryonic matter, and $\omega_\Lambda=-1$ for the Cosmological Constant (note that this is also the minimal physical value, because smaller values predict weird phenomena). However, since each contribution makes the others negligible for certain values of $a(t)$, the thermal history of the Universe is usually divided into different `domination' eras during which only the relevant fluid is considered. For example, near the singularity only stiff matter and radiation are relevant, while observations lead us to believe that today's Universe is going through a Cosmological Constant era. This case of negative pressure is rather interesting, and will now be expanded upon.\newline

\subsection{The Cosmological Constant Problem}
The acceleration equation has been written in \eqref{F2std(a)}. From this equation, one can deduce that the Universe decelerates during its expansion if $\rho + 3P > 0$, while it accelerates if $\rho + 3P < 0$.

The first condition seems to be always satisfied by ordinary fluids. Yet, in 1998, two independent groups, led by Riess \cite{art:universeexpansion1} and Perlmutter \cite{art:universeexpansion2}, showed that the Universe is actually accelerating during its expansion.

The discovery may lead to two different conclusions:
\begin{itemize}
\item{it might be wrong to use General Relativity because the dynamics is modified \cite{art:odintsov};}
\item{there could be a component of the Universe with a unusual equation of state, such as:
\begin{equation}
\label{eq:acccondition}
P < -\frac{1}{3}\rho
\end{equation}
which is currently dominating the Universe dynamics \cite{art:weinberg}.}
\end{itemize}
Without considering corrections to General Relativity, one can justify the measured acceleration of the Universe by adding a Cosmological Constant $\Lambda$ in the Einstein field equations:
\begin{equation}
G_{\mu\nu} - \Lambda g_{\mu\nu} = \chi T_{\mu\nu}
\end{equation}

If one shifts this new term to the right-hand side as proposed by Weinberg in \cite{art:weinberg}, it is interpreted as a physical phenomenon and not as a bare property of space-time. The field equations become then:
\begin{equation}
\label{eq:einsteincorrecteddx}
G_{\mu\nu} = \chi (T_{\mu\nu} + \rho_\Lambda g_{\mu\nu})
\end{equation}
where $\rho_\Lambda = \frac{\Lambda}{\chi}$ is the energy density related to the Cosmological Constant, and it is such that:
\begin{subequations}
\begin{equation}
T_\Lambda^{\mu\nu} = -P g^{\mu\nu} = \rho_\Lambda g^{\mu\nu}
\end{equation}
and from the continuity equation \eqref{continuity(a)} we obtain:
\begin{equation}
\dot \rho_\Lambda = 0
\end{equation}
\end{subequations}
and thus $\rho_\Lambda$ is a constant energy density.

Before the first evidence of acceleration of the Universe, the strong observational upper bound, i.e. $\Lambda < 10^{-120}$, led many particle physicists to suspect some fundamental principle to exist, in order to have $\Lambda=0$, as discussed in \cite{art:barrow}. Such a principle does not exist and the attempt to set the Cosmological Constant to zero has failed.
 
Many tried to explain the presence of a constant energy density as the result of the vacuum state of the quantum fields that fill the Universe. This approach leads for the scalar field to:
\begin{equation}
\label{eq:filedapprho}
\rho_\Lambda = \frac{k_\text{max}^4}{16\pi^2}
\end{equation}
where $k_\text{max}$ is the momentum cut-off, i.e. the energy scale at which the theory is believed to lose its validity. It can be estimated as the Planck energy, being widely believed that Planck length is the scale at which both gravitational and quantum effects need to be simultaneously taken into account. With this assumption, $\rho_\Lambda$ is $10^{122}$ times bigger than the energy density related to the Cosmological Constant measured today: "This is probably the worst theoretical prediction in the history of physics!" (from \cite{book:citlambda}).

\subsection{Hamiltonian formulation and the volume variable}
We will now restate the dynamics of the FLRW Universe in the framework of the Hamiltonian formulation of gravity, using a new variable $V=a^3$ that we will refer to as `volume'. The reason for this choice, as we will see in the next sections, is that this is the only variable with which the polymer parameter and the critical density are independent on the scale factor \cite{art:claudia}. 

The ADM line element in the homogeneous and isotropic model becomes \cite{book:primordialcosmology}:
\begin{equation}
ds^2=N^2(t)\,c^2dt^2-a^2(t)\big[dr^2+r^2 d\theta^2+r^2 \sin^2{\theta} d\varphi^2\big]
\end{equation}
Due to this, in the presence of an energy density $\rho=\rho(a)$ and substituting the RW metric directly in the Einstein-Hilbert action, the following Hamiltonian constraint is obtained:
\begin{equation}
\ham_\text{FLRW}(a,p_a)=-\frac{\chi}{24\pi^2}\frac{p_a^2}{a}+2\pi^2\rho a^3=0
\label{generalHam(a)}
\end{equation}
where $p_a=-\frac{12\pi^2}{\chi N}a\dot{a}$ is the momentum conjugate to $a$.

By performing a canonical transformation, we rewrite the Hamiltonian constraint as function of the volume variable and its momentum conjugate $P_V$:
\begin{equation}
\ham_\text{FLRW}(V,P_V)=-\frac{B}{2}P_V^2 V+2\pi^2\rho V=0
\label{Hamstd}
\end{equation}
with $B=\frac{3\chi}{4\pi^2}$. The independent cosmological equations are now derived from the Hamilton equation $\dot{V}=\pdv{\ham_\text{FLRW}}{P_V}$:
\begin{gather}
H^2=\frac{1}{9}\frac{\dot{V}^2}{V^2}=\frac{\chi}{3}\rho
\label{F1std}
\\
\dot{\rho}=-\frac{\dot{V}}{V}\left[\rho-\pdv{(\rho V)}{V}\right]
\end{gather}
The former is exactly the same as equation \eqref{F1std(a)}, while the latter coincides with the continuity equation \eqref{continuity(a)} under the identification $-\pdv{(\rho V)}{V}=P$ and will have solution $\rho(V)=\rho_0V^{-(1+\omega)}$.

Substituting the density as a function of $V$ in the Hamiltonian constraint, the system is analytically solvable and the evolution of the volume in time is given by:
\begin{equation}
V(t)=\left[\pi(1+\omega)\sqrt{(B\rho_0)}\,t\right]^{\frac{2}{1+w}}
\label{V(t)std}
\end{equation}

Explicitating all the constants and using Planck units $\rho_{P}=t_{P}^{-4}$,  $\tau=\frac{t}{t_{P}}$ with $t_{P}=\sqrt{G}$, we can rewrite our equations as dimensionless:
\begin{gather}
H_\text{ad}^2(Q)=H^2t_P^2=\frac{8\pi}{3}Q,\quad Q=\frac{\rho}{\rho_P},
\\
H_\text{ad}^2(V)=\frac{8\pi}{3}V^{-(1+\omega)},\quad\rho_0=\rho_P
\\
V(\tau)=\left[(1+\omega)\sqrt{6\pi\bar{Q}}\, \tau\right]^{\frac{2}{1+\omega}},\quad \bar{Q}=\frac{\rho_0}{\rho_P}
\end{gather}
The plots are shown in following sections (Figs.  \ref{H2polygraph} and \ref{V(t)polygraph}).

\section{Polymer Quantum Mechanics} \label{polymer}
Polymer representation is an alternative representation of Quantum Mechanics, non-unitarily connected to the standard Schrödinger representation. The introduction of a fundamental area in Loop Quantum Gravity (LQG)
leads to a bounce, i.e. a minimum of the scale factor, removing the singularity (see \cite{art:lqg1,art:lqg2,art:lqg3}). In analogy to LQG, polymer representation introduces a fundamental scale in the Hilbert space. When applied to Cosmology, it leads to the appearance of a bounce for the volume of the Universe.

Following \cite{art:corichi}, we now introduce the polymer representation of Quantum Mechanics.

Given the orthonormal basis $\ket {\mu_i}$ for the Hilbert space $\mathcal{H}'$, where $\mu_i \in \mathbb{R}$, $i=1,...,N$ and such that $\Braket{\mu_i|\mu_j}=\delta_{i,j}$, the Hilbert space $\mathcal{H}_\text{poly}$ for the polymer representation is built by the completion of $\mathcal{H}'$. In such a space we can define two fundamental operators:
\begin{subequations}
\begin{equation}
\hat \epsilon \ket \mu = \mu \ket \mu
\end{equation}
\begin{equation}
\hat s(\lambda) \ket \mu =\ket {\mu+\lambda}
\end{equation}
\end{subequations}
respectively label and shift operators. $\hat s(\lambda)$ is a family of parameter-dependent unitary operators. Yet, they are discontinuous and, therefore, they cannot be generated by the exponentiation of a self-adjoint operator.

Let us now consider a Hamiltonian system with canonical variables $q$ and $p$. In the momentum polarization, a state $\ket \psi$ has wave function $\psi(p)=\Braket{p|\psi}$, and then, for the fundamental states, we have:
\begin{equation}
\psi_\mu (p) = \Braket{p|\mu}=e^{i\mu p}
\end{equation}

Defining the multiplication operator $\hat V(\lambda)$ by:
\begin{equation}
\hat V(\lambda) \psi_\mu(p) = e^{i\lambda p} e^{i\mu p} = \psi_{\mu+\lambda}(p)
\end{equation}
we see that $\hat V(\lambda)$ is the shift operator in $\mathcal{H}_\text{poly}$ and it is clear that the momentum operator $\hat p$ cannot exist as the generator of translations. On the other hand, as regards the coordinate operator $\hat q$, it can be defined as the following differential operator:
\begin{equation}
\hat q \psi_\mu (p) = -i\frac{\partial}{\partial p} \psi_\mu (p) = \mu \psi_\mu(p)
\end{equation}
and it is the label operator in $\mathcal{H}_\text{poly}$.

It is possible to prove that the Hilbert space of the wave functions in such a polarization is given by $\mathcal{H}_\text{poly}=L^2(\mathbb{R}_B,d\mu_H)$, where $\mathbb{R}_B$ is the Bohr compactification of the real axis and $d\mu_H$ is the Haar measure.

If one wants so study the coordinate polarization, in which $\psi(q) = \Braket{q|\psi}$, it is possible to see that the fundamental wave functions are Kroenecker deltas. In this case it is the translation operator to be discontinuous which again implies the non-existence of the momentum operator and it can be proved that the Hilbert space is $\mathcal{H}_\text{poly}=L^2(\mathbb{R}_d, d\mu_c)$, where $\mathbb{R}_d$ is the real axis with discrete topology and $d\mu_c$ is the counting measure.

Given the impossibility of well defining both $\hat q$ and $\hat p$, the dynamics cannot be directly implemented. For this reason, we have to approximate the momentum operator by defining a regular graph $\gamma_{\mu_0}=\{q \in \mathbb{R} : q=n\mu_0 \text{ with } n \in \mathbb{Z}\}$, where $\mu_0$ is the fundamental scale introduced by the polymer representation. Consequently, defining $\mu_n=n\mu_0$, we consider the subspace $\mathcal{H}_{\gamma_{\mu_0}} \subset \mathcal{H}_\text{poly}$ which contains all those states $\ket \psi$ such that:
\begin{equation}
\ket \psi = \sum_n b_n \ket{\mu_n}
\end{equation}
where $\sum_n |b_n|^2 < \infty$. Now the translation operator acts only by discrete steps in order to remain on $\gamma_{\mu_0}$:
\begin{equation}
\hat V(\mu_0) \ket {\mu_n} = \ket {\mu_{n+1}}
\end{equation}

When the condition $p \ll \frac{1}{\mu_0}$ is satisfied, we can write:
\begin{equation}
\label{eq:ppoly}
p \approx \frac{1}{\mu_0} \sin{(\mu_0 p)} = \frac{1}{2i\mu_0}\bigl(e^{i\mu_0 p}-e^{-i\mu_0 p} \bigr)
\end{equation}
and in return we can approximate the action of the momentum operator by that of $\hat V(\mu_0)$:
\begin{equation}
\begin{aligned}
\hat p_{\mu_0} \ket{\mu_n} &= \frac{1}{2i\mu_0} \bigl(\hat V(\mu_0) - \hat V(-\mu_0) \bigr) \ket{\mu_n} = \\
&= \frac{i}{2\mu_0} \bigl(\ket{\mu_{n+1}} - \ket{\mu_{n-1}} \bigr) 
\end{aligned}
\end{equation}

As regards the squared momentum operator, we can choose two different approximations:
\begin{subequations}
\begin{equation}
\label{eq:p2oppol}
p^2 \approx \frac{2}{\mu^2_0} \bigl(1- \cos{(\mu_0 p)} \bigr)
\end{equation}
so that:
\begin{equation}
\label{eq:pmu01shift}
\hat p^2_{\mu_0} \ket {\mu_n} = \frac{1}{\mu_0^2} \bigl(2-\hat V(\mu_0) -\hat V(-\mu_0) \bigr) \ket{\mu_n}
\end{equation}
\end{subequations}
or
\begin{subequations}
\begin{equation}
p^2 \approx \frac{1}{\mu^2_0} \sin^2{(\mu_0 p)}
\end{equation}
so that:
\begin{equation}
\label{eq:pmu02shift}
\hat p^2_{\mu_0} \ket {\mu_n} = \frac{1}{4\mu_0^2}\bigl(2-\hat V(2\mu_0)-\hat V(-2\mu_0) \bigr) \ket{\mu_n}
\end{equation}
\end{subequations}
which are equivalent through a reparametrization of the polymer scale.

Now, we can implement a Hamiltonian operator on the graph:
\begin{equation}
\hat H_{\gamma_{\mu_0}} = \frac{1}{2m} \hat p^2_{\mu_0} + \hat V(\hat q)
\end{equation}
where $\hat V(\hat q)$ is the potential.

If one wants to quantize a system using the momentum polarization of the polymer representation, one has to approximate the momentum operator using eq. \eqref{eq:pmu01shift} or eq. \eqref{eq:pmu02shift}, while the coordinate operator is a derivative operator, given by:
\begin{equation}
\label{eq:qoppol}
\hat q \psi(p) = i \frac{\partial}{\partial p} \psi(p)
\end{equation}

Alternatively, one can work with semiclassical states, i.e. states peaked in their classical value, by operating the following substitution on the classical hamiltonian:
\begin{subequations}
\begin{equation}
p \rightarrow \frac{1}{\mu_0} \sin{(\mu_0 p)}
\end{equation}
\begin{equation}
\label{eq:p2poly}
p^2 \rightarrow \frac{2}{\mu^2_0} \bigl(1- \cos{(\mu_0 p)} \bigr) \quad \text{or} \quad p^2 \rightarrow \frac{1}{\mu^2_0} \sin^2{(\mu_0 p)}.
\end{equation}
\end{subequations}

\section{Semiclassical Polymer Evolution of the Universe} \label{Gabriele}
We will now apply some features of Polymer Quantum Mechanics to the Hamiltonian formulation of the Standard Cosmological Model. In the spirit of the Ehrenfest theorem \cite{art:claudia}, this will be done at a semiclassical level, meaning that we will not develop a full quantum theory but will apply quantum modifications to the classical evolution.

\subsection{Exact approach} \label{sec:exact}
The first modification, called \emph{exact substitution}, consists in using the approximation of $p^2$ as a squared sine \eqref{eq:p2poly} in the FLRW Hamiltonian, and deriving the dynamics through the standard Hamilton equations. This has already been done in \cite{art:claudia}, and therefore we will only report the main results.

\begin{figure}[hbt!]
\centering
\includegraphics[scale=0.55]{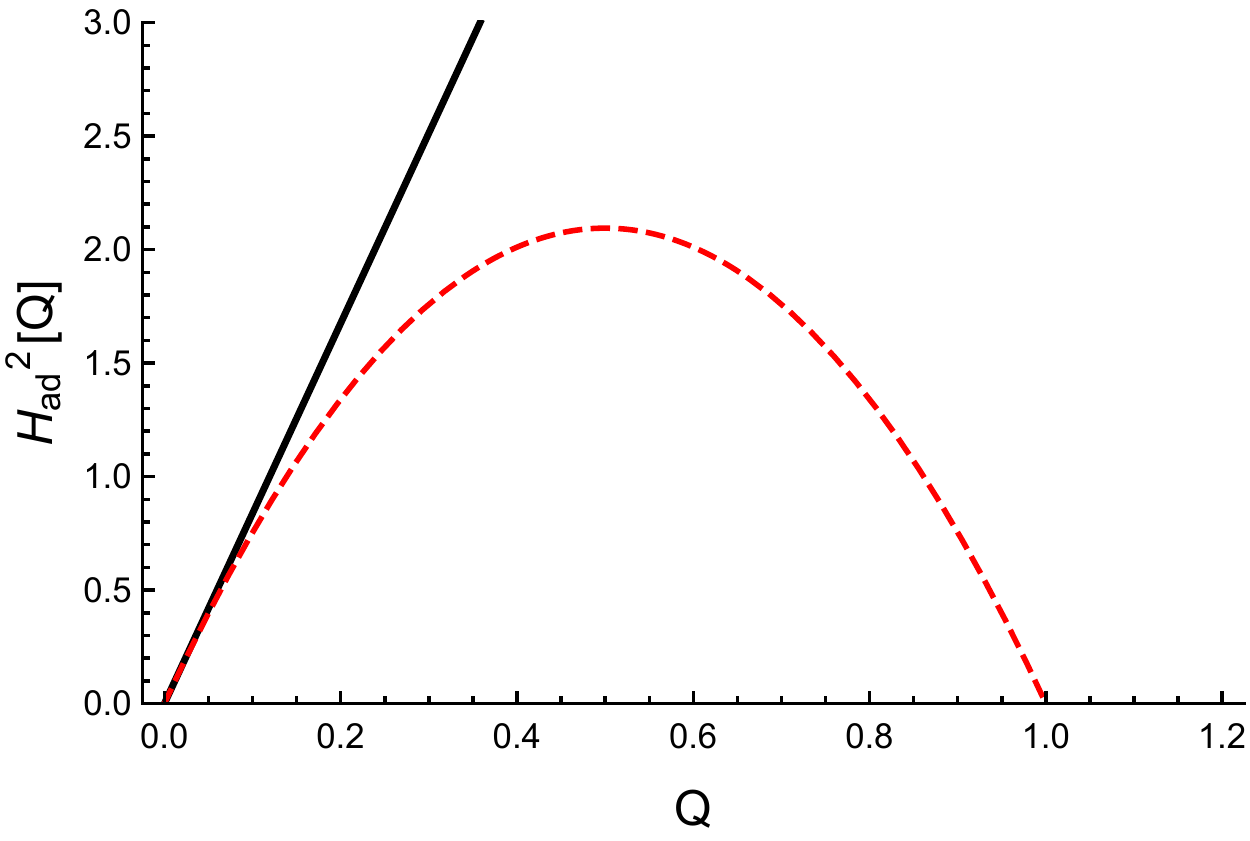}
\quad
\includegraphics[scale=0.55]{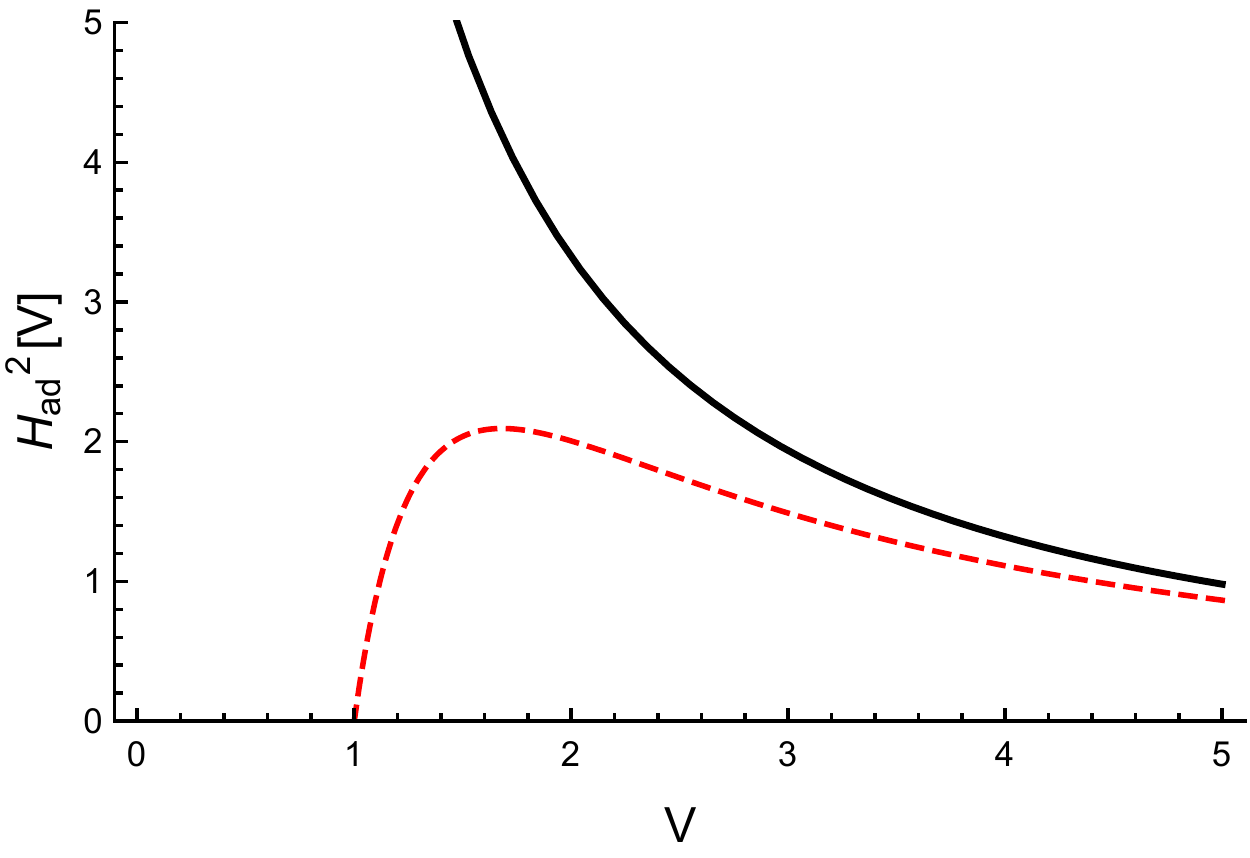}
\caption{The confrontation between the classical (continuous) and polymer (dashed) dimensionless Hubble parameter as function of density $Q$ (above) and of volume $V$ (below) for $\omega=1/3$; for the polymer functions the parameters are $Q_\mu=1$ and $\mu_0=\sqrt{\frac{3}{2\pi^3}}$.}
\label{H2polygraph}
\end{figure}

\begin{figure}[hbt!]
\centering
\includegraphics[scale=0.55]{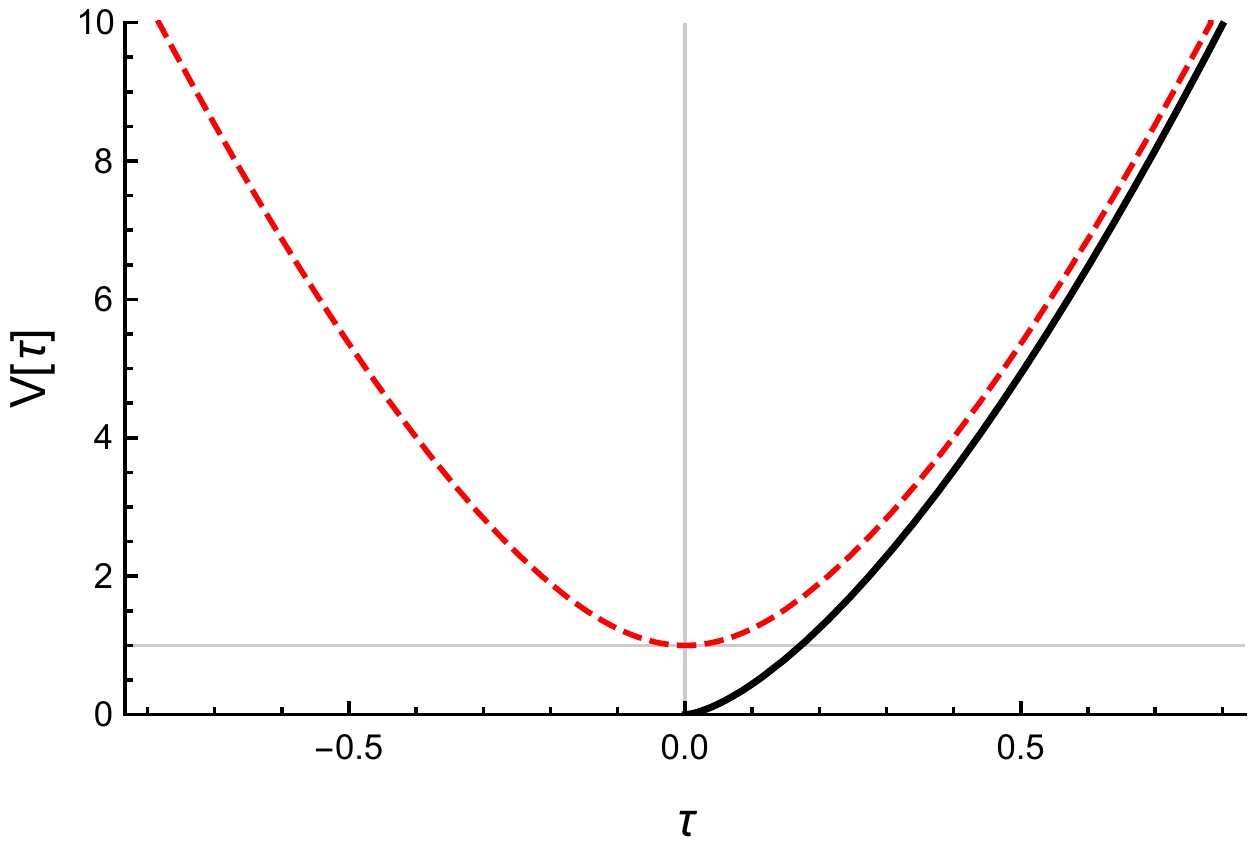}
\caption{The confrontation between the classical (continuous) and polymer (dashed) dimensionless volume as function of $\tau$ with $\mu_0=\sqrt{\frac{3}{2\pi^3}}$ for $\omega=1/3$. The minimal volume is highlighted.}
\label{V(t)polygraph}
\end{figure}

The Hamiltonian function becomes:
\begin{equation}
\ham_\text{FLRW}^\text{poly}=-\frac{B}{2\mu_0^2}\Big[\sin[2](\mu_0 P_V)\Big]V+2\pi^2\bar{\rho}V^{-\omega}=0
\end{equation}
Through the Hamilton equations, we find the modified first Friedmann equation and the volume evolution:
\begin{gather}
H^2=\frac{1}{9}\frac{\dot{V}^2}{V^2}=\frac{\chi}{3}\rho\left(1-\frac{\rho}{\rho_{\mu}}\right),\quad\rho_{\mu}=\frac{B}{4\pi^2}\frac{1}{\mu_0^2}
\label{polyF1}
\\
V(t)=\left(\frac{4\pi^2\rho_0}{B}\right)^{\frac{1}{1+\omega}}\bigg[\frac{B^2}{4}(1+\omega)^2t^2+\mu_0^2\bigg]^{\frac{1}{1+\omega}}
\label{V(t)poly}
\end{gather}
First of all, by taking the classical limit $\mu_0\rightarrow0$ and therefore $\rho_\mu\rightarrow\infty$, the standard equations \eqref{F1std} and \eqref{V(t)std} are recovered. This can be better visualized by rewriting these equations in their dimensionless form, with the same procedure used for the classical case:
\begin{gather}
H_\text{ad}^2(Q)=H^2t_P^2=\frac{8\pi}{3}Q\left(1-\frac{Q}{Q_\mu}\right),\quad Q_\mu=\frac{\rho_\mu}{\rho_P}
\\
H_\text{ad}^2(V)=\frac{8\pi}{3}V^{-(1+\omega)}\bigg(1-V^{-(1+\omega)}\bigg)
\label{H2adpoly(V)}
\\
V(\tau)=\left(\left[(1+\omega)\sqrt{6\pi}\,\tau\right]^2+\frac{2\pi^3}{3}\mu_0^2\right)^{\frac{1}{1+\omega}}
\end{gather}
It's clear how, since $H^2$ can now be zero for a finite value of the density $\rho=\rho_\mu$, the evolution of the volume as function of time will have a critical point, and already from equation \eqref{V(t)poly} we see a non zero minimum for the volume. This is shown in Figs. \ref{H2polygraph} and \ref{V(t)polygraph}, where the polymer-modified evolution is compared to the classical one: we see that, while for low energies (i.e. great volumes and times) the latter is recovered, for high energies the change is substantial in that $H^2$ goes to zero for a finite value of the density and of the volume, and the volume itself reaches a non zero minimum and starts to increase again. So, this deformation, already at a semiclassical level, results in a Big Bounce scenario and effectively solves the singularity. The minimal volume is easily calculated to be $V_0=\left(\frac{\rho_0}{\rho_{\mu}}\right)^{\frac{1}{1+\omega}}$, which becomes $V_0=\left(\frac{2\pi^3}{3}\mu_0^2\right)^{\frac{1}{1+\omega}}$ in its dimensionless form. In equation \eqref{H2adpoly(V)} we put $Q_\mu=1$, which is equivalent to asking that $V_0=1$; this automatically fixes the value of the (dimensionless) polymer parameter to $\mu_0=\sqrt{\frac{3}{2\pi^3}}\approx0.22$, which corresponds to a polymer lattice parameter of $L_\text{poly}=\sqrt[3]{\mu_0}\,\ell_P\approx0.60\,\ell_P$. Of course this is just an estimate, and it depends on the definition of Planck density, but being of the order of the Plank length it is well within the expectations.
\newline
\paragraph*{The sign of the Cosmological Constant}
Looking at eq. \eqref{polyF1}, it is possible to have a negative Cosmological Constant.

It is interesting to consider a Universe filled with some kind of energy density $\rho_s$ associated to a source, and a Cosmological Constant energy density $\rho_\Lambda$. The Friedmann equation is then:
\begin{equation}
\label{eq:lambdaneg}
H^2 = \frac{\chi}{3} (\rho_s + \rho_{\Lambda}) \biggl( 1 - \frac{\rho_s + \rho_{\Lambda}}{\rho_\mu} \biggr)
\end{equation}

The condition:
\begin{equation}
\label{eq:Hpos}
H^2 \ge 0
\end{equation}
has to be always satisfied. It implies:
\begin{equation}
\label{eq:conditionsign}
(\rho_s + \rho_{\Lambda}) (\rho_\mu -\rho_s - \rho_{\Lambda}) \ge 0 
\end{equation}

Condition \eqref{eq:conditionsign} is verified in two cases:
\begin{subequations}
\begin{equation}
\label{eq:firstcase}
\begin{cases}
\rho_s \ge -\rho_{\Lambda} \\
\rho_s \le \rho_\mu - \rho_{\Lambda}
\end{cases}
\end{equation}
or
\begin{equation}
\label{eq:secondcase}
\begin{cases}
\rho_s \le -\rho_{\Lambda} \\
\rho_s \ge \rho_\mu - \rho_{\Lambda}
\end{cases}
\end{equation}
\end{subequations}

\begin{figure}
\centering
\includegraphics[scale=0.55]{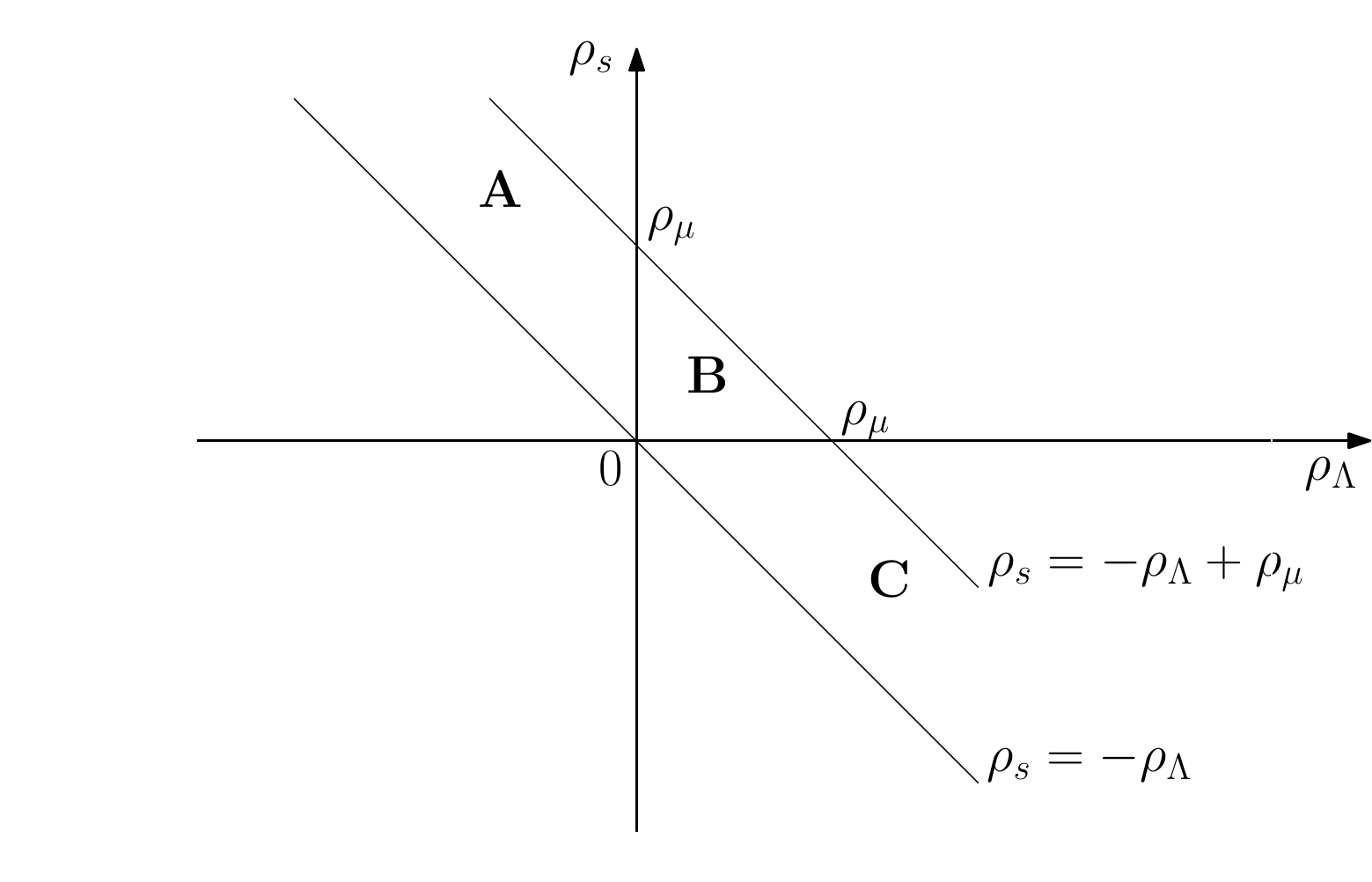}
\caption[Study of the sign of the Cosmological Constant on $(\rho_{\Lambda};\,\rho_s)$ plane]{Study of the sign of the Cosmological Constant on $(\rho_{\Lambda};\,\rho_s)$ plane. The regions in the plotting are the ones that satisfy condition \eqref{eq:Hpos}.}\label{fig:signlambda}
\end{figure}

While \eqref{eq:secondcase} has no intersection on $(\rho_{\Lambda};\,\rho_s)$ plane, \eqref{eq:firstcase} identifies three different regions where condition \eqref{eq:Hpos} is satisfied, as shown in Fig. \ref{fig:signlambda}.

In region A, i.e. $(\rho_{\Lambda} \le 0) \,\land\, (-\rho_{\Lambda} \le \rho_s \le -\rho_{\Lambda}+\rho_\mu)$, the energy density related to the Cosmological Constant is negative but is balanced by matter density; in region B, i.e. $(0<\rho_{\Lambda}<\rho_\mu) \,\land\, (0 \le \rho_s \le -\rho_{\Lambda}+\rho_\mu)$, both the densities are positive; in region C, i.e. $[(0<\rho_{\Lambda}<\rho_\mu) \,\land\, (-\rho_{\Lambda}\le \rho_s \le 0)] \lor [(\rho_{\Lambda} > \rho_\mu) \,\land\, (-\rho_{\Lambda} \le \rho_s \le -\rho_{\Lambda}+\rho_\mu)]$, the matter density is negative and then this region is not of physical interest.

In conclusion, we have found a region where it is possible to have a negative value for the Cosmological Constant.

\subsection{Perturbative approach}
Polymer features can be implemented in the cosmological model also through another approach. If we consider the Schr\"odinger commutation relation $\comm{q}{p}=i$ of standard Quantum Mechanics and use the polymer approximation \eqref{eq:ppoly}, we obtain a cosine that can be expanded in a power series:
\begin{equation}
\begin{gathered}
\comm{q}{p}\rightarrow\comm{q}{\frac{\sin(\mu_0 p)}{\mu_0}}=\cos(\mu_0 p)\comm{q}{p}=
\\
=i\cos(\mu_0 p)\approx i\left(1-\frac{\mu_0^2p^2}{2}\right)
\end{gathered}
\end{equation}
At a semiclassical level this modified commutation rule becomes a rule for Poisson brackets. The polymer-modified evolution can then be derived from the standard unmodified Hamiltonian constraint \eqref{Hamstd} through the scheme $\pb{V}{P_V}=(1-\frac{\mu_0^2}{2}P_V^2)$.

This scheme is made to look similar to the so-called Generalized Uncertainty Principle (GUP). This approach states that by modifying the canonical commutation relations (CCR) of standard Quantum Mechanics, it is possible to obtain the Generalized Uncertainty Principle that was derived in the low energy limit of String Theory \cite{art:kempf}:
\begin{gather}
\comm{q}{p}=i(1+\lambda p^2)
\\
\Delta q\,\Delta p\geq\frac{1}{2}\Big(1+\lambda(\Delta p)^2+\lambda\expval{\hat{p}}^2\Big)
\end{gather}
where $\lambda$ is the GUP parameter. This principle implies a fundamental minimum uncertainty on position $\Delta q_0=\sqrt{\lambda}$, and therefore through the simple modification of the CCR it is possible to implement string features in Quantum Mechanics without going too deep in String Theory, and to reproduce a low energy limit of Brane Cosmology \cite{art:brane1,art:brane2,art:brane3}. Notice how the GUP modification of the CCR, under the identification $\lambda\leftrightarrow\frac{\mu_0^2}{2}$, coincides with the polymer deformation of Poisson brackets apart from a sign. During this section we will study the `perturbative' polymer approach and present in parallel the results that would be obtained with a GUP approach, i.e. by deriving the dynamics thorough the modified commutation relation used as a modified rule for Poisson brackets. For a more detailed mathematical confrontation between the two approaches, see \cite{art:gorji}.

The first Friedmann equation in the perturbative polymer approach takes the form:
\begin{equation}
H^2=\frac{1}{9}\frac{\dot{V}^2}{V^2}=\frac{\chi}{3}\rho\left(1-\frac{\rho}{2\rho_{\mu}}\right)^2
\label{F1polypoisson}
\end{equation}

\begin{figure}[hbt!]
\centering
\includegraphics[scale=0.55]{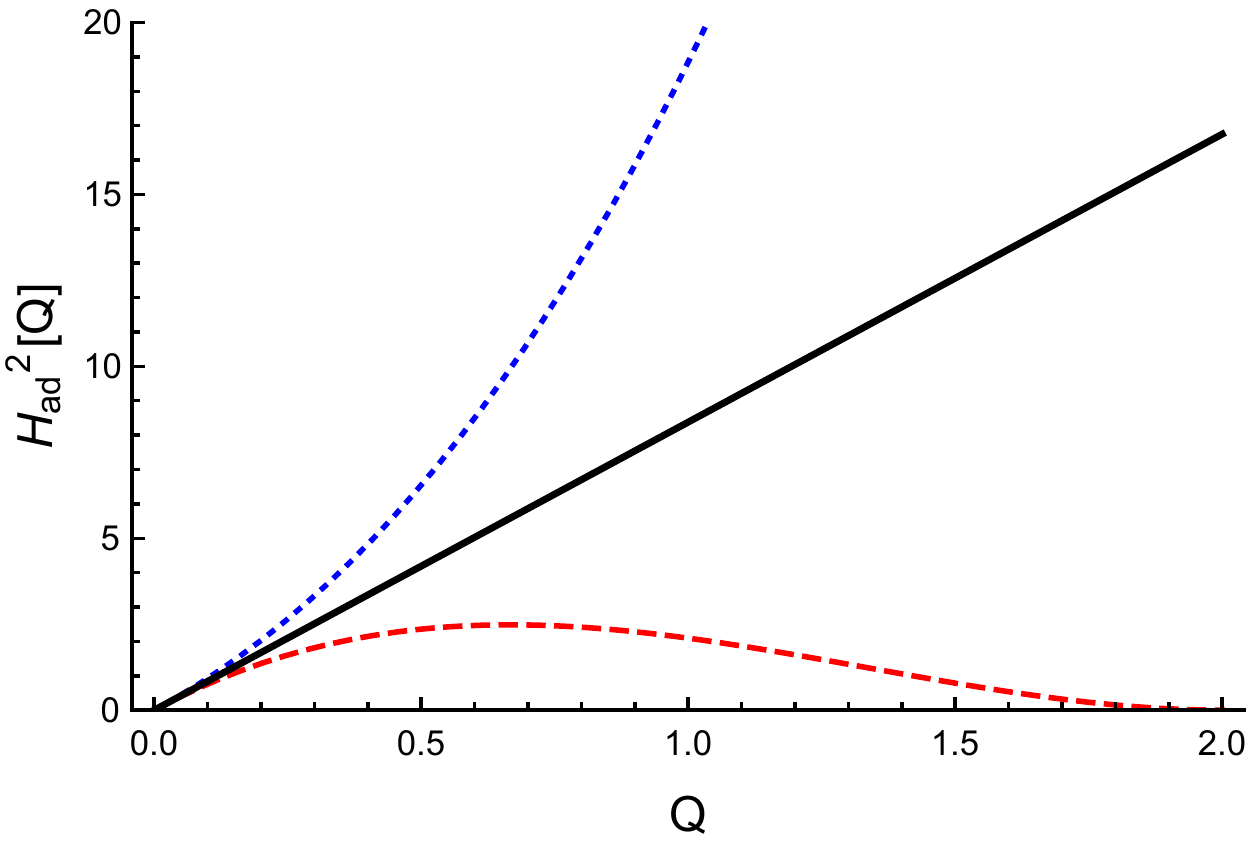}
\quad
\includegraphics[scale=0.55]{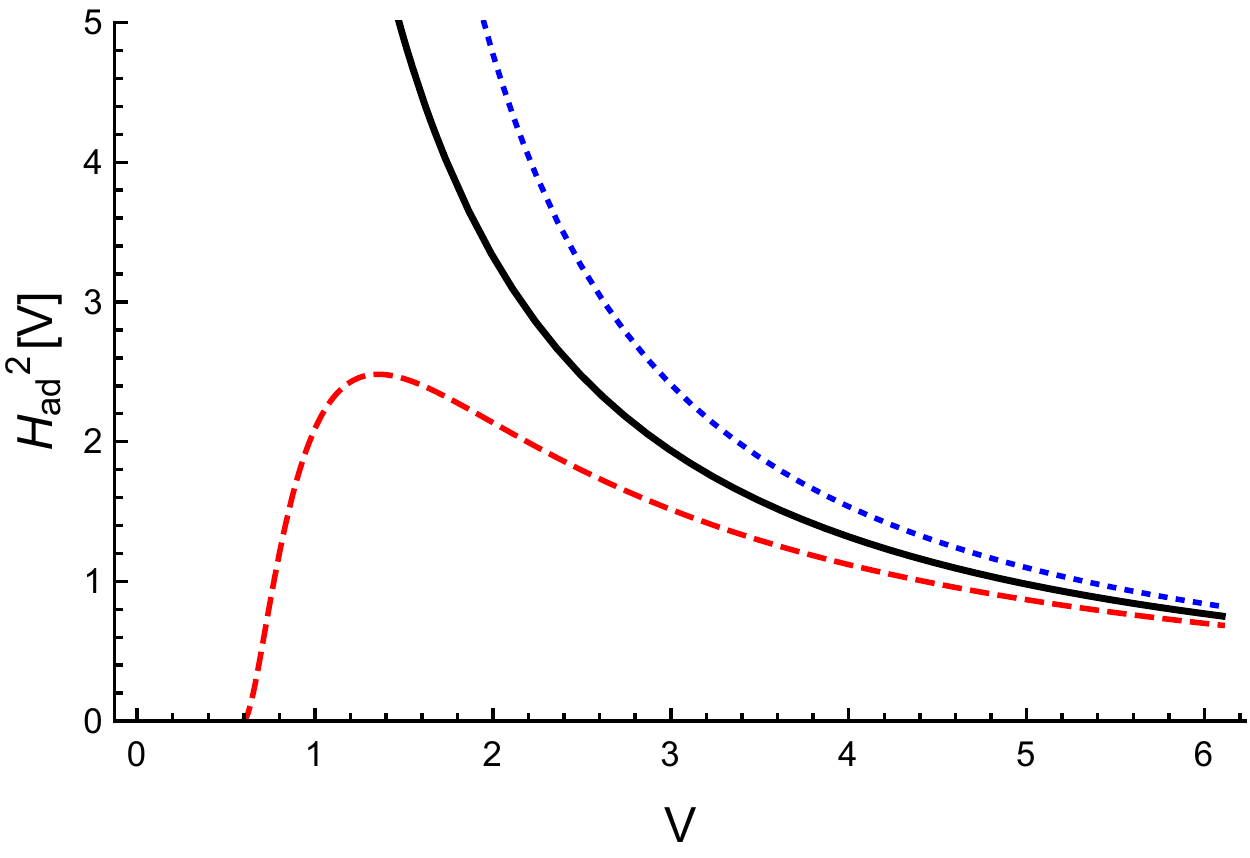}
\caption{The confrontation between the classical (continuous), perturbative polymer (dashed) and GUP (dotted) dimensionless Hubble parameter as function of density $Q$ (above) and of volume $V$ (below) for $\omega=1/3$; for the polymer and GUP functions the parameters are $Q_i=1$ and $\lambda=\frac{\mu_0^2}{2}=\frac{3}{4\pi^3}$.}
\label{H2GUPvsPolygraph}
\end{figure}

The equivalent equation in the GUP framework is $H^2=\frac{\chi}{3}\rho\left(1+\frac{\rho}{2\rho_{\lambda}}\right)^2$ with $\rho_\lambda=\frac{B}{8\pi^2}\frac{1}{\lambda}$. Notice how the different sign completely changes the dynamics: now $H^2$ doesn't become zero for a finite value of the density, and therefore in this model a Bounce is not possible. Thus, we conclude that the GUP semiclassical deformation doesn't solve the singularity. Figure \ref{H2GUPvsPolygraph} shows the comparison between the two approaches, with the equations rewritten in dimensionless form: $H_\text{ad}^2=\frac{8\pi}{3}Q\left(1\pm\frac{Q}{2Q_i}\right)^2=\frac{8\pi}{3}V^{-(1+\omega)}\left(1\pm\frac{V^{-(1+\omega)}}{2}\right)^2$, where the subscript $i$ indicates both $\mu$ and $\lambda$ and in the last equation we put $Q_i=1$.

Now, equation \eqref{F1polypoisson} is slightly different from the exact polymer substitution case, in that $H^2$ goes to zero for $\rho=2\rho_{\mu}$ instead of $\rho=\rho_{\mu}$. However the last factor is squared, and if we expand it we obtain:
\begin{equation}
\begin{gathered}
H^2=\frac{\chi}{3}\rho\Bigg[1-\frac{2\rho}{2\rho_{\mu}}+\frac{\rho^2}{4\rho_{\mu}^2}\Bigg]\approx
\\
\approx\frac{\chi}{3}\rho\left[1-\frac{\rho}{\rho_{\mu}}+\order{\frac{\rho^2}{\rho_{\mu}^2}}\right]
\end{gathered}
\end{equation}
We see that the exact equation \eqref{polyF1} is recovered in the limit $\rho\ll\rho_{\mu}$; already here we can say that this approach is a low energy limit of the exact one.

\begin{figure}[hbt!]
\centering
\includegraphics[scale=0.55]{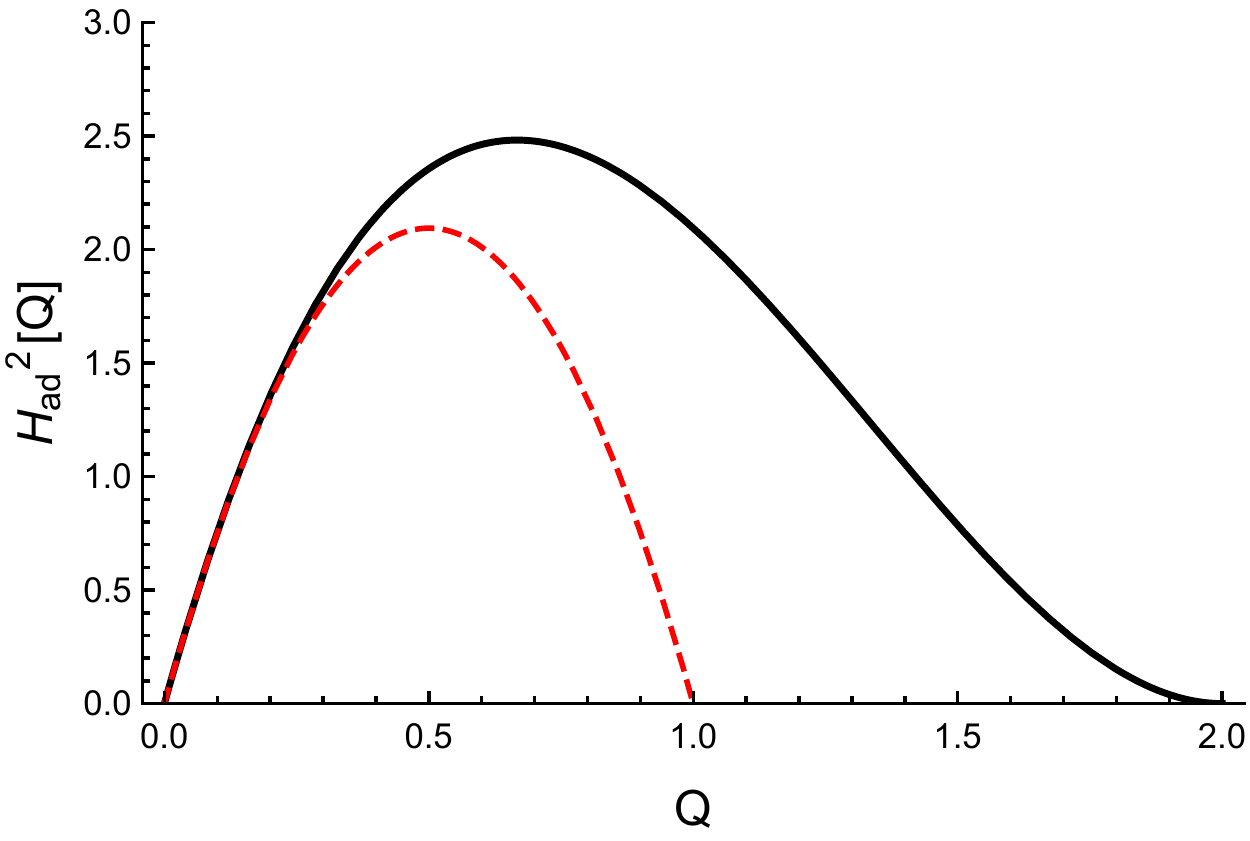}
\quad
\includegraphics[scale=0.55]{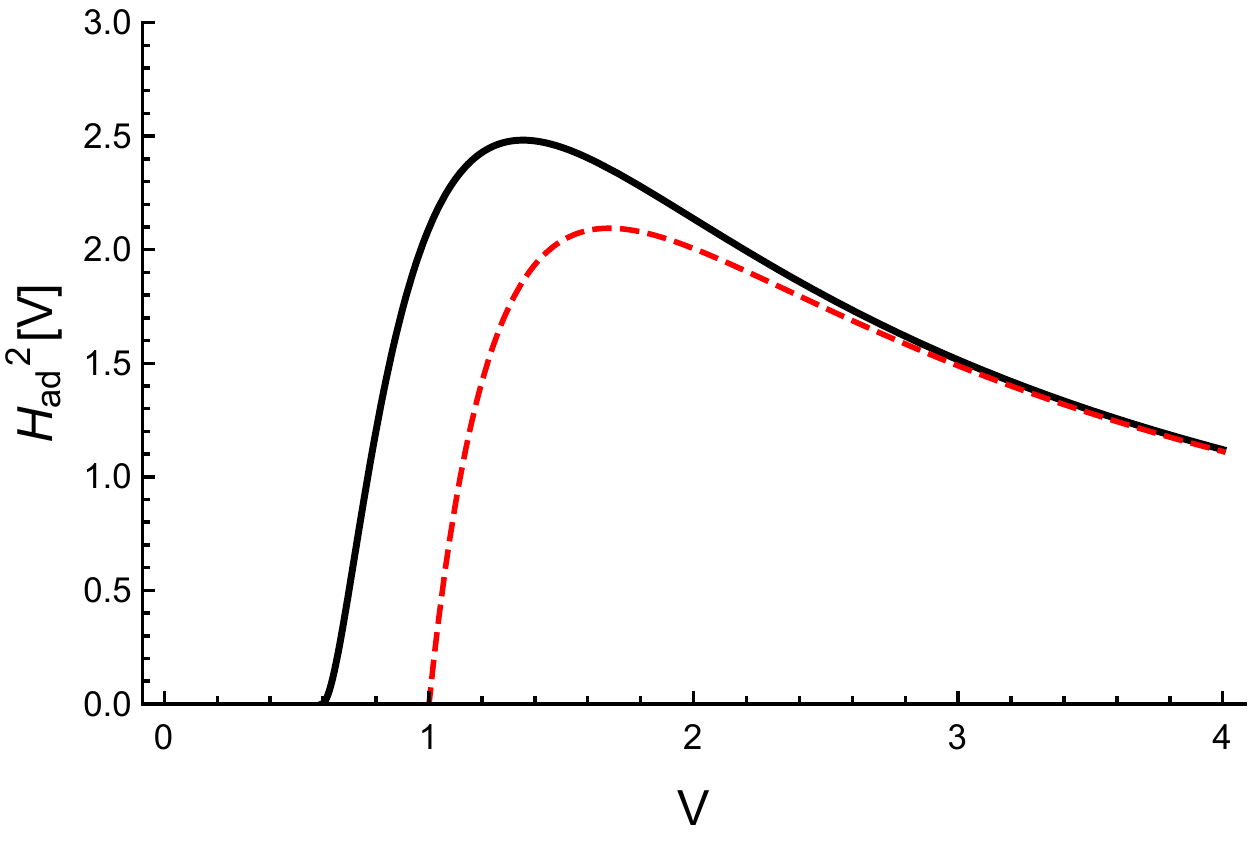}
\caption{The confrontation between the perturbative (continuous) and exact polymer (dashed) dimensionless Hubble parameter as function of density $Q$ (above) and of volume $V$ (below) for $\omega=1/3$; the parameters are $Q_\mu=1$ and $\mu_0=\sqrt{\frac{3}{2\pi^3}}$.}
\label{H2polypoissongraph}
\end{figure}
The dimensionless form of the new first Friedmann equation is 
\begin{gather}
H_\text{ad}^2(Q)=\frac{8\pi}{3}Q\left(1-\frac{Q}{2Q_\mu}\right)^2
\\
H_\text{ad}^2(V)=\frac{8\pi}{3}V^{-(1+\omega)}\left(1-\frac{V^{-(1+\omega)}}{2}\right)^2
\end{gather}
where in the latter we put $Q_\mu=1$ and $\bar{\rho}=\rho_P$ as before. In Fig. \ref{H2polypoissongraph} we see the comparison between these behaviours and the ones coming from the exact approach. We can see how for low energies they coincide exactly, while for high energies, where the $Q^2$ term becomes relevant and is not negligible anymore, they are slightly different. In particular, the values of maximal density and minimal volume are different, as mentioned before.

Solving the evolution leads to the following implicit expression for the volume variable as function of time:
\begin{equation}
\frac{V^{\frac{1+\omega}{2}}}{A_1\mu_0}+\frac{1}{2}\ln\Bigg[1-A_1^2\mu_0^2V^{-(1+\omega)}\Bigg]=\pm\frac{B}{\sqrt{2}\,\mu_0}(1+\omega)\,t
\end{equation}
where $A_1^2=\frac{2\pi^2\rho_0}{B}$. This is rewritten as dimensionless, plotted and compared with its counterpart from the exact approach.
\begin{equation}
V^{\frac{1+\omega}{2}}+\frac{A_0}{2}\ln\left[1-A_0^2V^{-(1+\omega)}\right]=\pm(1+\omega)\sqrt{6\pi}\,\tau
\end{equation}
with $A_0^2=\frac{\pi^3}{3}\mu_0^2$.
\begin{figure}[hbt!]
\centering
\includegraphics[scale=0.5]{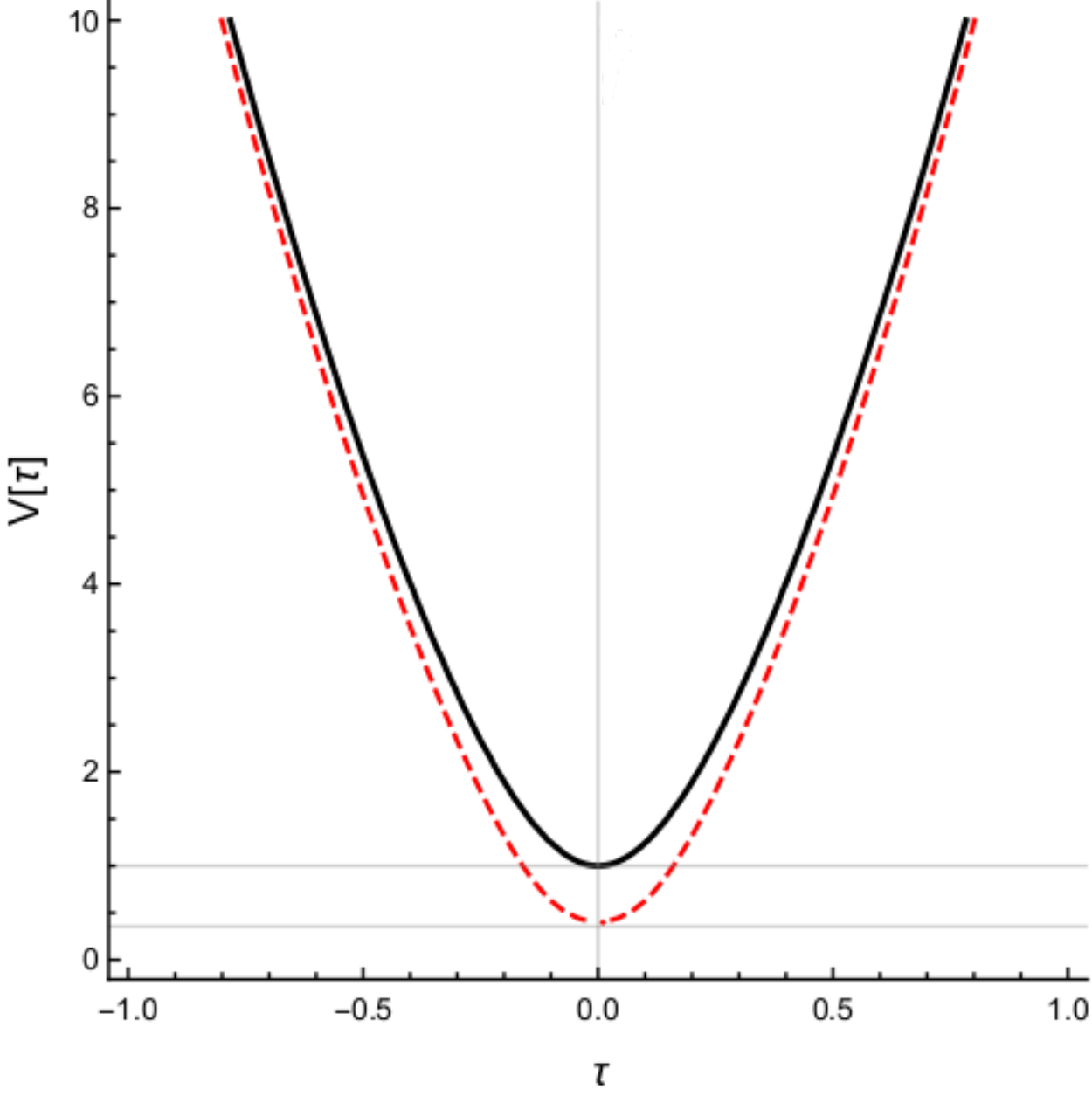}
\caption{The confrontation between the exact (continuous) and perturbative polymer (dashed) dimensionless volume as function of $\tau$ for $\omega=1/3$ with $\mu_0=\sqrt{\frac{3}{2\pi^3}}$. The minimal volumes are highlighted.}
\label{V(t)polypoissongraph}
\end{figure}
As we can see from Fig. \ref{V(t)polypoissongraph}, the difference between the two is mainly the value of the minimal volume, that in this case becomes $V_0=\left(\frac{\rho_P}{2\rho_\mu}\right)^{\frac{1}{1+\omega}}=\left(\frac{\pi^3}{3}\mu_0^2\right)^{\frac{1}{1+\omega}}$, but the fundamental character of this model being a Bounce Cosmology is unaltered. \\
We would like to stress that higher orders of expansion in the polymer parameter are expected to produce no significant new physics on the Bounce Cosmology. In fact, the considered non-commutative formulation of Polymer Quantum Mechanics must converge on the exact polymer representation of the cosmological dynamics, actually equivalent to LQC. On the other hand, the situation for the interesting analysis in \cite{art:cianfrani} is different. Here, the standard approach discussed in \cite{art:lqg2,art:lqg3} receives corrections from the graph structure underlying the space representation, in particular from small terms in the inverse node number. This revised Loop Cosmology predicts pre-Bounce oscillations of the Universe scale factor and its main conceptual merit is to reproduce the Bounce morphology of LQC on a more well-grounded representation of the space graph.

\begin{figure}[hbt!]
\centering
\includegraphics[scale=0.5]{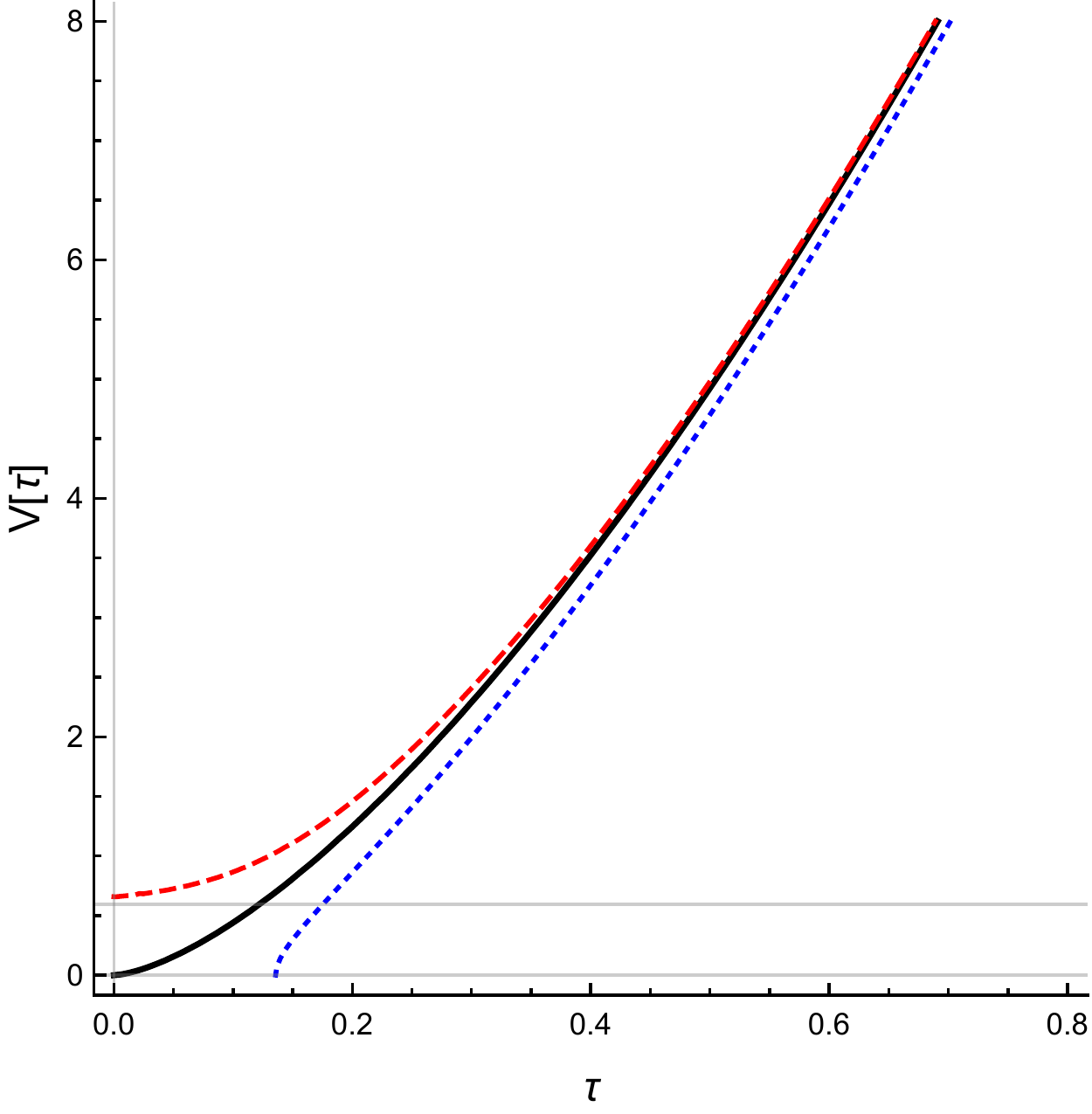}
\caption{The confrontation between the classical (continuous), perturbative polymer (dashed) and GUP (dotted) dimensionless volume as function of $\tau$ for $\omega=1/3$ and $\lambda=\frac{\mu_0^2}{2}=\frac{3}{4\pi^3}$.}
\label{V(t)GUPvsPolygraph}
\end{figure}

In the GUP approach the following implicit equation for the volume as function of time is obtained:
\begin{gather}
\frac{V^{\frac{1+\omega}{2}}}{C_1}-\arctan\left(C_1\,V^{-\frac{1+\omega}{2}}\right)=\frac{B}{\sqrt{\lambda}}\frac{(1+\omega)}{2}\,t
\\
V^{\frac{1+\omega}{2}}+C_0\arctan(C_0\,V^{\frac{1+\omega}{2}})=(1+\omega)\sqrt{6\pi}\,\tau
\end{gather}
with $C_1=2\pi\sqrt{\frac{\lambda\rho_0}{B}}$ and $C_0=\sqrt{\frac{4\pi^3}{3}\lambda}$. Fig. \ref{V(t)GUPvsPolygraph} compares the evolution in the two approaches. As we can see, in the GUP framework the volume $V$ still goes to zero and the singularity is still present.

In conclusion, we can confirm that while the polymer deformation solves the singularity by introducing a Big Bounce, the GUP deformation does not. Besides, we can conclude that our perturbative approach is the low-energy approximation of the polymer exact approach, i.e. it is to Loop Quantum Cosmology what the Generalized Uncertainty Principle is to Brane Cosmology. \\

\section{Vacuum Energy Density of the scalar field in polymer representation} \label{Paolo}
In this section we study the vacuum state of the scalar field in Polymer Quantum Mechanics. We derive the energy spectrum of the harmonic oscillator in polymer representation, in order to evaluate the vacuum energy density of the massless scalar field in a flat FLRW Universe.

\subsection{Energy spectrum of the harmonic oscillator in polymer representation of Quantum Mechanics}
The Hamiltonian function of the harmonic oscillator $\hat H = \frac{\hat p^2}{2m} + \frac{1}{2}m\omega^2 \hat q^2$, due to the substitutions \eqref{eq:qoppol} and \eqref{eq:p2oppol}, leads to the polymer Hamiltonian $\hat H_\mu$:
\begin{equation}
\label{eq:oscillatorHmu}
\hat H_\mu = \frac{1}{m\mu^2}\bigr[1- \cos{(\mu p)}\bigl] -\frac{1}{2}m\omega^2\partial^2_p
\end{equation}

A state of energy $E$, in momentum polarization, is described by the wave function $\psi(p)$. Thus, the following Schrödinger equation, in polymer representation, can be studied:
\begin{equation}
\label{eq:oscillatorpolymerschrodinger}
\hat H_\mu \psi(p) = E \psi(p)
\end{equation}

Through an opportune reparametrization of the polymer scale $\mu$ and the following variable change:
\begin{subequations}
\begin{equation}
\label{eq:mathieuu}
u = \mu p + \frac{\pi}{2}
\end{equation}
\begin{equation}
\partial_p = \mu \partial_u
\end{equation}
\end{subequations}

eq. \eqref{eq:oscillatorHmu} turns into:
\begin{equation}
\label{eq:oscillatorpolymerschrodingermathieu}
\partial^2_u \psi(u) + \biggl[ \frac{2E}{\omega g} - \frac{\hbar^2}{2g^2} -\frac{\hbar^2}{2g^2} \cos{(2u)} \biggr] \psi(u) = 0
\end{equation}
where we have defined:
\begin{equation}
\label{eq:g1}
g=m\omega \mu^2
\end{equation}
that is a dimensionless parameter that measures the intensity of polymer corrections. Equation \eqref{eq:oscillatorpolymerschrodingermathieu} is the Mathieu equation that leads to the following even and odd solutions:
\begin{subequations}
\begin{equation}
\label{eq:polevenpsi}
\psi_{2n}(u) = \pi^{-1/2}\, \text{ce}_n \biggl( \frac{1}{4g^2}, u \biggr)
\end{equation}
\begin{equation}
\label{eq:polevenE}
E_{2n} = \omega \biggl[ \frac{1}{4g} + \frac{g}{2}A_n\biggl(\frac{1}{4g^2}\biggr) \biggr]
\end{equation}
\end{subequations}
and
\begin{subequations}
\begin{equation}
\label{eq:poloddpsi}
\psi_{2n+1}(u) = \pi^{-1/2}\, \text{se}_{n+1} \biggl( \frac{1}{4g^2}, u \biggr)\end{equation}
\begin{equation}
\label{eq:poloddE}
E_{2n+1} = \omega \biggl[ \frac{1}{4g} + \frac{g}{2}B_{n+1}\biggl(\frac{1}{4g^2}\biggr) \biggr]
\end{equation}
\end{subequations}
where $A_n$ and $B_n$ are the Mathieu characteristic value functions, $\text{ce}_n$ and $\text{se}_n$ are, respectively, the elliptic cosine and sine of order $n$.

\begin{figure}
\centering
\includegraphics[scale=0.6]{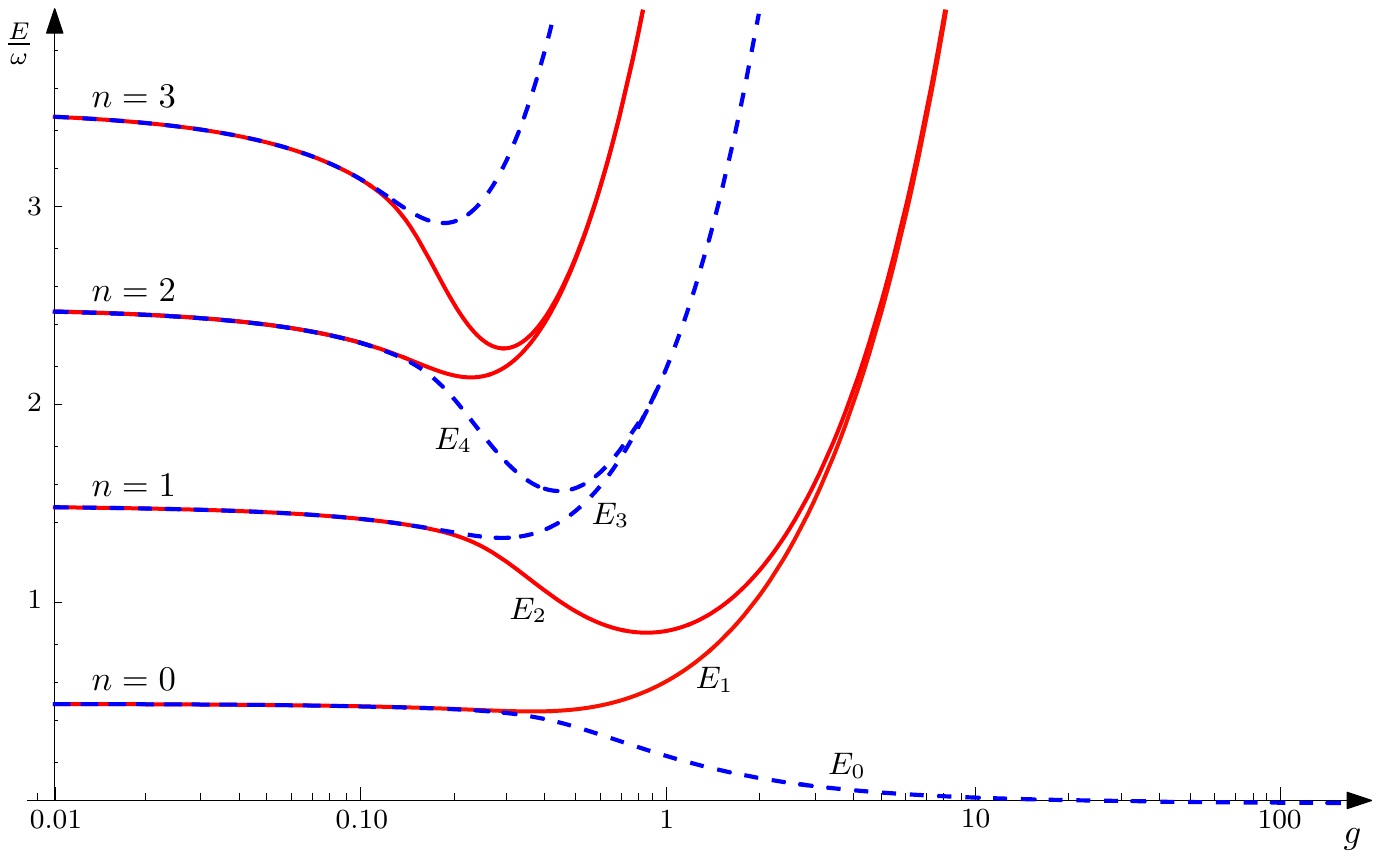}
\caption[vv]{Plotting of  $\frac{E}{\omega}$ vs. $g$. In this plot, the first states ($n=0,1,2,3$) of the energy spectrum of a polymer quantum harmonic oscillator are shown. They are $\pi$-periodic (dashed blue lines) and $\pi$-antiperiodic (continuous red lines).}\label{fig:polyenergy}
\end{figure}

In Fig. \ref{fig:polyenergy}, the plotting of $\frac{E}{\omega}$ vs. $g$ for the fundamental state and some of the first excited states of polymer quantum harmonic oscillator is shown.

The energy spectrum is degenerate both in the range of small polymer corrections (small $g$) and of big polymer corrections (big $g$). The fundamental state is the only one with an energy that does not diverge for $g \rightarrow \infty$.

As discussed in \cite{book:handbook}, $\psi_{2n}(u)$ are $\pi$-periodic for even $n$ and $\pi$-antiperiodic for odd $n$, while $\psi_{2n+1}(u)$ are $\pi$-antiperiodic for even $n$ and $\pi$-periodic for odd $n$, as shown in Fig. \ref{fig:polyenergy}.

The polymer representation should reproduce the standard Schrödinger quantization in the limit $\mu\rightarrow 0$. In light of this consideration, it is easy to show that eqs. \eqref{eq:polevenE} and \eqref{eq:poloddE} tend to the energy spectrum of the standard harmonic oscillator. Using the asymptotic expansion from \cite{book:handbook}, in the limit of small polymer corrections, i.e. for $g \rightarrow 0$, we get:
\begin{equation}
E_{2n} \simeq E_{2n+1} \simeq \omega \biggl[ n + \frac{1}{2} - \frac{(2n+1)^2 + 1}{16} g \biggr]
\end{equation}

In the opposite limit, i.e. for $g \rightarrow \infty$, that is the limit of big polymer correction to standard quantization, it can be shown that:
\begin{equation}
\label{eq:polE0}
E_0 \simeq \frac{\omega}{4g} \rightarrow 0
\end{equation}
which is the only value of $n$ for which the energy eigenvalue does not diverge.

Also in polymer representation, the fundamental state is the one with $n=0$. Yet, $n$ has no physical interpretation, due to the impossibility to define annihilation and creation operators in polymer representation.

As regards the fundamental state, it is possible to show that eq. \eqref{eq:polevenpsi} reproduces correctly the standard fundamental state, in the case of $n=0$ and $\mu \rightarrow 0$. In \cite{art:mathieuexp} and \cite{art:mathieuexp0}, asymptotic expansions of periodic solutions of the Mathieu equation have been studied. Using those expressions, one has:
\begin{equation}
\label{eq:ceexpansion}
\text{ce}_0(q,u) = D_0(\alpha) + o({q^{-\frac{1}{2}}})
\end{equation}
being
\begin{subequations}
\begin{equation}
\alpha = 2q^{\frac{1}{8}} \cos{u}
\end{equation}
\begin{equation}
D_m(\alpha) = \frac{1}{2^{m/2}} e^{-\frac{\alpha^2}{4}} H_m \biggr(\frac{\alpha}{\sqrt{2}}\biggl)
\end{equation}
\end{subequations}
where $H_m$ is the Hermite polynomial and $q$ is a parameter that appears in the standard formulation of Mathieu equation.

It can be easily shown that, in the limit $g \rightarrow 0$:
\begin{equation}
\text{ce}_0(q,u) \sim e^{-\frac{p^2}{2\hbar m \omega}}
\end{equation}
that is exactly the fundamental state of a standard harmonic oscillator in momentum representation.

\subsection{Evaluation of the vacuum energy density}
Aim of this section is to study the scalar field in the background of an isotropic and homogeneous expanding flat Universe. In particular, we are interested in the evaluation of the energy density for the fundamental state of the field, i.e. the vacuum energy density.

Given the metric \eqref{RWmetric}, the action of the theory is:
\begin{equation}
\label{eq:fieldfrwaction}
S_\phi = \int d^4x\sqrt{-g}\, \frac{1}{2}g^{\mu\nu}\partial_\mu \phi \partial_\nu \phi
\end{equation}

The metric tensor can be split as follows:
\begin{equation}
g_{\mu\nu} = \text{diag}\bigl(g_{00}, q_{ab} \bigr)
\end{equation}
where $q_{ab}$ is the spatial metric, and, then, $\sqrt{-g}=\sqrt{-q}=a^3(t)$

We can write the lagrangian of the field as follows:
\begin{equation}
\label{eq:frwlagrangian}
\mathscr{L}_\phi = \sqrt{-q} \biggl[\frac{1}{2} \dot \phi ^2 + q^{ab} \partial_a \phi \partial_b \phi \biggr]
\end{equation}
and, consequently, the conjugate momentum density is:
\begin{equation}
\label{eq:frwmomentum}
\Pi = \frac{\partial \mathscr{L}_\phi}{\partial \dot \phi} = \sqrt{-q} \dot \phi
\end{equation}

and due to the lagrangian \eqref{eq:frwlagrangian} and the momentum \eqref{eq:frwmomentum}, we can derive the Hamiltonian function and density, the latter of which defined as $\mathcal{H} \equiv \Pi \dot \phi - \mathscr{L}_\phi$:
\begin{subequations}
\begin{equation}
\label{eq:frwhphi}
H_\phi = \int d^3 \bar x \mathcal{H}
\end{equation}
\begin{equation}
\label{eq:frwHamiltoniandensity}
\mathcal{H} = \frac {\Pi^2}{2 \sqrt{-q}} - \frac{1}{2} \sqrt{-q} q^{ab} \partial_a \phi \partial_b \phi
\end{equation}
\end{subequations}

It is convenient to change coordinates:
\begin{subequations}
\begin{equation}
(t,x^i) \rightarrow (t, \bar x^i)
\end{equation}
\begin{equation}
\label{eq:frwchange}
\bar x^i = a(t) x^i
\end{equation}
\end{subequations}
and then the line element \eqref{RWmetric} becomes:
\begin{equation}
ds^2 = dt^2 - (d\bar x^2 +d\bar y^2 + d\bar z^2)
\end{equation}

After this change, the Fourier decomposition of the field $\bar \phi$ and the momentum $\bar\Pi$ is analogous to the one developed in the Minkowski background, as in \cite{art:propagator}. One reduces the Universe in a fiducial box of finite volume $\bar V = \int d^3 \bar x$ and then:
\begin{subequations}
\label{eq:frwdecomposition}
\begin{equation}
\bar \phi (t, \bar {\underline x}) = \frac{1}{\sqrt{\bar V}} \sum_{\bar{\underline k}} \bar \phi_{\bar k}(t) e^{i \bar {\underline k} \cdot \bar {\underline x} }
\end{equation}
\begin{equation}
\bar \Pi (t, \bar {\underline x}) = \frac{1}{\sqrt{\bar V}} \sum_{\bar{\underline k}} \bar \uppi_{\bar k}(t) e^{i \bar {\underline k} \cdot \bar {\underline x} }
\end{equation}
\end{subequations}
where $\bar{\underline x}$ is the 3-vector $(\bar x, \bar y, \bar z)$, $\bar {\underline k}$ is the Fourier mode in the new coordinates and it is $\bar {\underline k} = \frac{\underline k}{a(t)}$, being $\underline k$ the comoving frequency. Moreover, $\phi_k$ and $\uppi_k$ have been chosen as real functions and they have the dimension, respectively, of $length^{1/2}$ and $length^{-1/2}$.

It is easy to define the $\delta$ functions:
\begin{subequations}
\label{eq:frwdelta}
\begin{equation}
\delta_{\underline {\bar k}, \underline {\bar k'}} = \frac{1}{\bar V} \int d^3 \bar x\, e^{i (\underline {\bar k} - \underline {\bar k'}) \cdot \bar{\underline x}}
\end{equation}
\begin{equation}
\delta^3_{\underline {\bar x}, \underline {\bar y}} = \frac{1}{\bar V} \sum_{\underline{\bar k}} e^{i (\underline {\bar x} - \underline {\bar y}) \cdot \bar{\underline k}}
\end{equation}
\end{subequations}

Using these definitions in the Hamiltonian \eqref{eq:frwhphi}, we can rewrite it as a composition of independent harmonic oscillators: 
\begin{subequations}
\label{eq:hdeckbar}
\begin{equation}
H_\phi(t) = \sum_{\underline {\bar k}} \mathcal{H}_{\bar k}(t)
\end{equation}
\begin{equation}
\label{eq:htimek}
\mathcal{H}_{\bar k}(t) = \frac{1}{2}\bar \uppi_{\bar k}^2 + \frac{1}{2} \bar k^2 \bar \phi^2_{\bar k}
\end{equation}
\end{subequations}
where $\bar k^2=\frac{k^2}{a^2}$ is the norm of the 3-vector $\underline {\bar k}$.

We would like to express this result in the old coordinates $(t,x,y,z)$. In order to do so, the understanding of how the Fourier components $\bar \phi_{\bar k}$ and $\bar \uppi_{\bar k}$ are related to the comoving components $\phi_k$ and $\uppi_k$ is needed.

When $x \rightarrow \bar x = ax$, the transformation for the field and the momentum is:
\begin{subequations}
\label{eq:riparametrizzazioneax}
\begin{equation}
\bar \phi (t, \underline {\bar x}) = \phi (t, \underline x)
\end{equation}
\begin{equation}
\bar \Pi (t, \underline {\bar x}) = \frac{\Pi (t, \underline x)}{a^3}
\end{equation}
\end{subequations}
because the field is scalar while the momentum is actually a scalar density.

Imposing \eqref{eq:riparametrizzazioneax} in \eqref{eq:frwdecomposition}, we obtain a relation for the Fourier components of the field and its momentum for the transformation $x \rightarrow \bar x = ax$:
\begin{subequations}
\label{eq:riparametrizzazioneaxmodi}
\begin{equation}
\bar \phi_{\bar k} = a^{3/2} \phi_k
\end{equation}
\begin{equation}
\bar \uppi_{\bar k} = \frac{\uppi_k}{a^{3/2}}
\end{equation}
\end{subequations}

In terms of the old coordinates, we get:
\begin{subequations}
\begin{equation}
\label{eq:frwhdec}
H_\phi (t) = \sum_{\underline k} \mathcal{H}_k(t)
\end{equation}
\begin{equation}
\label{eq:frwhkbar}
\mathcal{H}_k(t) = \frac{\uppi_k^2}{2a^3}+\frac{1}{2} \frac{k^2}{a^2}a^3 \phi^2_k
\end{equation}
\end{subequations}
where, as already discussed, the 3-vector $\underline k$ is comoving and the cosmological expansion related time dependence is in the $a(t)$ terms. This result has been found in \cite{art:coreani} and \cite{art:primordialpolymerperturbations}.

Due to the presence of the scale factor $a(t)$, the Hamiltonian is time dependent. In Quantum Mechanics, a physical state described by the wave function $\psi$ is a solution of the Schrödinger equation:
\begin{equation}
\label{eq:timeschr}
H(t) \psi = i\partial_t \psi
\end{equation}

If the Hamiltonian is time dependent, its eigenstates do not satisfy \eqref{eq:timeschr}, hence they are not physical states. Nevertheless, in this work, the words 'eigenstate', 'eigenvalue' and 'fundamental state' will be used in relation to the time dependent Hamiltonian.

The polymer quantization of the single Fourier mode is now implemented through the substitutions \eqref{eq:qoppol} and \eqref{eq:p2oppol}:
\begin{subequations}
\label{eq:frwpolvar}
\begin{equation}
\label{eq:frwpolphik}
\phi_k \rightarrow i\partial_{\uppi_k}
\end{equation}
\begin{equation}
\label{eq:frwpolpik}
\uppi_k \rightarrow \frac{a^{3/2}}{\mu}\sin{\biggl( \frac{\mu \uppi_k}{a^{3/2}} \biggr)}
\end{equation}
\end{subequations}
where the factor $a^{3/2}$ has been included in order to have $\hat V(\mu)$ transforming as a scalar during the expansion of the Universe.

It is essential to say that the physical results that we obtain here are closely linked to eq. \eqref{eq:frwpolpik}. This choice of $\uppi_k$ in polymer representation is not unique and it has been inspired by \cite{art:primordialpolymerperturbations,art:nonsingular}. This new ambiguity deserves attention and further investigation.

The polymer Hamiltonian operator becomes:
\begin{equation}
\label{eq:frwhkbarpol}
\mathcal{H}_k^\mu = \frac{1}{\mu^2} \biggl[ 1-\cos{\biggl( \frac{\mu \uppi_k}{a^{3/2}} \biggr) } \biggr] -\frac{1}{2} a^3 \frac{k^2}{a^2} \frac{\partial^2}{\partial \uppi_k^2}
\end{equation}

We solve the following equation, in order to evaluate the eigenvalues and eigenstates of the operator \eqref{eq:frwhkbarpol}:
\begin{equation}
\label{eq:frwpolsch}
\mathcal{H}_k^\mu (t) \psi_k(t,\uppi_k) = \mathcal{E}_k(t) \psi_k(t, \uppi_k)
\end{equation}

Solutions of such an equation are given by the results of the previous section, once we define:
\begin{subequations}
\begin{equation}
u\equiv \frac{\mu \uppi_k}{a^{3/2}} +\frac{\pi}{2}
\end{equation}
\begin{equation}
\label{eq:frwg}
g\equiv \bar k\mu^2=\frac{k}{a} \mu^2
\end{equation}
\end{subequations}
we obtain:
\begin{subequations}
\begin{equation}
\label{eq:frwpolevenpsi}
\psi_k^{2n}(t,u) = \pi^{-1/2}\, \text{ce}_n \biggl( \frac{1}{4g^2}, u \biggr)
\end{equation}
\begin{equation}
\label{eq:frwpolevenE}
\mathcal{E}_k^{2n}(t) = \frac{k}{a} \biggl[ \frac{1}{4g} + \frac{g}{2}A_n\biggl(\frac{1}{4g^2}\biggr) \biggr]
\end{equation}
\end{subequations}
and
\begin{subequations}
\begin{equation}
\label{eq:frwpoloddpsi}
\psi_k^{2n+1}(t,u) = \pi^{-1/2}\, \text{se}_{n+1} \biggl( \frac{1}{4g^2}, u \biggr)
\end{equation}
\begin{equation}
\label{eq:frwpoloddE}
\mathcal{E}_k^{2n+1}(t) = \frac{k}{a} \biggl[ \frac{1}{4g} + \frac{g}{2}B_{n+1}\biggl(\frac{1}{4g^2}\biggr) \biggr]
\end{equation}
\end{subequations}

The fundamental state is then:
\begin{subequations}
\label{eq:polfundamentalstatefrw}
\begin{equation}
\psi_k^{0}(t,u) = \pi^{-1/2}\, \text{ce}_0 \biggl( \frac{1}{4g^2}, u \biggr)
\end{equation}
\begin{equation}
\label{eq:frwfundamentalE}
\mathcal{E}_k^{0}(t) = \frac{k}{a} \biggl[ \frac{1}{4g} + \frac{g}{2}A_{0}\biggl(\frac{1}{4g^2}\biggr) \biggr]
\end{equation}
\end{subequations}

The vacuum state $\ket 0$ is defined as:
\begin{subequations}
\label{eq:vacuum}
\begin{equation}
\ket 0 = \prod_{\underline k} {\ket 0}_k
\end{equation}
\begin{equation}
\sum_{\underline k}{\mathcal{H}_k^\mu (t)} \ket 0 = \sum_{\underline k} \mathcal{E}_k^{0}(t) \ket 0 =E_0(t) \ket 0
\end{equation}
\end{subequations}
and we define the energy density of the vacuum as:
\begin{equation}
\label{eq:frwrholambdadef}
\rho_\Lambda (t) \equiv \frac{E_0(t)}{\bar V}
\end{equation}

By means of eq. \eqref{eq:frwfundamentalE}, the energy density of the vacuum becomes:
\begin{equation}
\rho_\Lambda (t) = \frac{1}{\bar V} \sum_{\underline{\bar k}} \bar k \biggl[ \frac{1}{4g} + \frac{g}{2} A_0\biggl( \frac{1}{4g^2}\biggr) \biggr]
\end{equation}
where, for convenience, the sum is expressed in terms of $\underline{\bar k}$. It is important to remember that there is a factor $a^{-1}$ in the definition \eqref{eq:frwg} of $g$, and then $g$ is thought as a function of $\underline{\bar k}$.

Using the prescription $\frac{1}{\bar V} \sum_{\underline {\bar k}} \rightarrow \frac{1}{(2\pi)^3} \int d^3 \bar k$ to develop the continuum limit and switching to the variable $\underline k$, the energy density of the vacuum can be evaluated over all space:
\begin{subequations}
\begin{equation}
\rho_\Lambda (t) = \int \frac{d^3 k}{(2\pi)^3 a^3}\, \biggl[ \frac{1}{4\mu(k)^2} + \frac{k^2\mu^2(k)}{2a^2}A_0\biggl( \frac{a^2}{4k^2\mu^4} \biggr) \biggr] 
\end{equation}

Being the quantization of the single modes independent from one another, there is no a priori reason to ask for the polymer scale to be the same for each mode. From now on, we will consider a mode-dependent polymer scale $\mu(\underline k)$, which needs to satisfy the following condition:
\begin{equation}
\label{eq:prop}
\mu^2(\underline k) k \rightarrow 0, \qquad \text{for } k \rightarrow 0
\end{equation}
in order to correctly reproduce the standard propagator of the scalar field, as shown in \cite{art:propagator}.

The dependence of the energy of the fundamental state for the single mode is all in the magnitude $k$ and not in the direction or versus of the vector $\underline k$. Thus, it seems reasonable to choose $\mu(\underline k) \equiv \mu(k)$. We choose the following polymer scale:
\begin{equation}
\label{eq:mukfrw}
\mu(\bar k) \equiv \alpha \ell_P^{\frac{5}{2}}\bar k^2 = \alpha \ell_P^{5\over 2}\frac{k^2}{a^2}
\end{equation}
\end{subequations}
where $\ell_P$ is the Planck length and $\alpha$ is a dimensionless constant. This choice is the simplest polymer scale that makes the integral above converge. It satisfies the condition \eqref{eq:prop} and it has the correct dimension of $length^{1/2}$.

Defining the adimensional parameter $q=\alpha^{2/5} \ell_P k a^{-1}$ and noticing that the dependence from the frequency is all in $k$ and then $\int d^3k = 4\pi \int_0^\infty k^2 dk$, the energy density of the vacuum is in the end calculated:
\begin{equation}
\label{eq:frwdensity}
\rho_\Lambda = \frac{I_M}{16\pi^2\alpha^{\frac{8}{5}}\ell_P^4}
\end{equation}
where $I_M$ is the adimensional integral defined by:
\begin{equation}
\label{eq:im}
I_M \equiv \int_0^\infty dq\, \biggl[ 2 q^{-2} + 4 q^8 A_0 \biggl( \frac{1}{4}q^{-10} \biggr) \biggr] \simeq 2.776
\end{equation}

In the general definition \eqref{eq:frwrholambdadef}, $\rho_\Lambda$ is clearly a function of time. In fact, not only the Hamiltonian is explicitly time dependent, but the energy density has to be evaluated in the fiducial box of volume $\bar V = a^3 V_0$, being $V_0$ the comoving box. The volume of the box is a function of time due to the expansion of the Universe. These two different contributions do cancel in the evaluation of the energy density \eqref{eq:frwdensity}, which is a constant. This energy density does behave as an actual Cosmological Constant. Moreover, one can observe that the polymer scale \eqref{eq:mukfrw} is Planckian if $\alpha = O(1)$. In \cite{art:polbound}, an upper bound on polymer scale is discussed. Moreover, the polymer quantization is here implemented on the single Fourier mode and the coordinate variable is the Fourier component $\phi_k$ of the scalar field: there is no \emph{a priori} direct link between the polymer scale and a length scale in the physical space. If the polymer scale in the configuration space and lengths in the physical space were related, and if we asked $\alpha$ to make $\rho_\Lambda\simeq 10^{-120}$, i.e. the measured Cosmological Constant, the polymer scale would not be small, but comparable to the Hubble horizon. We would then observe macroscopical polymer corrections we clearly do not measure in the Universe we live in. The Cosmological Constant problem is then not solved by an opportune choice for the parameter $\alpha$.

\subsection{Cosmological dynamics with polymer scalar field}
Despite the time dependence of the Hamiltonian, it is worth studying what happens when we consider the vacuum energy density evaluated above inside the equations for the cosmological dynamics.

A modified Friedmann equation \eqref{polyF1} has been found. An effective energy density $\rho^{\text{eff}}$ has been introduced by the presence of the damping factor $1-\frac{\rho}{\rho_\mu}$, due to the polymer corrections to the semiclassical cosmological dynamics.
We now express the polymer scale $\mu_0$ as:
\begin{equation}
\label{eq:cosmomu}
\mu_0 = \beta {\ell_P}^2 \sqrt{\chi}
\end{equation}
where $\beta$ is an adimensional constant, $\ell_P$ is the Planck length and $\mu_0$ has the dimension of the inverse of $P_V$. Then, in the hypothesis that the energy density $\rho_\Lambda$ of the scalar field is the dominant component of the Universe, we explicitly evaluate the damping term that appears in the corrected Friedmann equation, by means of eq. \eqref{eq:frwdensity} and \eqref{polyF1}:
\begin{equation}
1-\frac{\rho_{\Lambda}}{\rho_\mu} = 1-\frac{\pi^2 I_M}{3} \frac{\beta^2}{\alpha^{8/5}}
\end{equation}
and one should ask this damping term to be $10^{-120}$ in order to solve the Cosmological Constant problem that has been presented. Despite this could be a mechanism for the reduction of the Cosmological Constant value, there is no physical reason for such a fine-tuning for the parameter $\alpha$, which is related to the polymer scale $\mu$ of the scalar field, and $\beta$, which is related to the polymer scale $\mu_0$ of the metric. In conclusion, the Cosmological Constant problem is not automatically solved, even considering polymer corrections to the cosmological dynamics.

\section{Propagation of gravitational waves on a flat  Semiclassical Polymer FLRW background} \label{Alberto}
In this section we investigate the propagation of gravitational waves when the cosmological background is described by a semiclassical polymer dynamics. We firstly recall the standard behaviour of cosmological gravitational waves and then study the effects induced by a modified Friedmann equation. 

\subsection{Gravitational waves on a classical flat FLRW background}
In order to study the propagation of gravitational waves on a classical FLRW background with null curvature \cite{book:weinberg}, we use perturbation theory by taking into account the following perturbed metric:
\begin{equation}
g_{\mu\nu}=\bar{g}_{\mu\nu}+h_{\mu\nu} \qquad |h_{\mu\nu}|\ll 1
\end{equation}
where $\bar{g}_{\mu\nu}$ is the metric tensor corrisponding to the line element \eqref{RWmetric} for the flat FLRW Universe on a synchronous reference frame. \\
We then write Einstein equations as:
\begin{equation}
R_{\mu\nu}=\chi S_{\mu\nu}=T_{\mu\nu}-\frac{1}{2}g_{\mu\nu}T^{\lambda}_{\lambda}
\end{equation}
so that the equation of propagation for gravitational waves is found by writing the first order linearized equations that will be indicated with the following notation:
\begin{equation}
\delta R_{\mu\nu}=\chi \delta S_{\mu\nu}
\end{equation}
At this point, a suitable infinitesimal transformation of coordinates can be chosen in order to have $h_{0i}=h_{00}=0$.\\
It is then necessary to write the linearized Christoffel symbols, $\Gamma^\alpha_{\beta\gamma}=\bar{\Gamma}^\alpha_{\beta\gamma}+\delta\Gamma^\alpha_{\beta\gamma}$, where $\delta\Gamma^\alpha_{\beta\gamma}$ is the Christoffel at the first order in $h_{\mu\nu}$. The only ones that survive are given by:
\begin{equation}
\delta\Gamma^0_{ij}=-\frac{1}{2}\dot{h}_{ij}
\end{equation}
\begin{equation}
\delta\Gamma^i_{jk}=-\frac{1}{2a^2}(h_{ij,k}+h_{ik,j}-h_{jk,i})
\end{equation}
\begin{equation}
\delta\Gamma^i_{0j}=-\frac{1}{2a^2}\Big(\dot{h}_{ij}-\frac{2\dot{a}}{a}h_{ij}\Big)
\end{equation}

It is then possible to write the linearized Ricci tensor $\delta R_{\mu\nu}$ whose non zero components are:
\begin{equation}
\delta R_{00}=\frac{1}{2a^2}\big(\ddot{h}_{ii}-\frac{2\dot{a}}{a}\dot{h}_{ii}+2\Big(\frac{\dot{a}^2}{a^2}-\frac{\ddot{a}}{a}\Big)h_{ii}\big)
\end{equation}
\begin{equation}
\delta R_{0i}=\frac{1}{2}\partial_t\Big(\frac{1}{a^2}(\partial_ih_{jj}-\partial_j h_{ij})\Big)
\end{equation}
\begin{equation}
\label{modi tensoriali}
\begin{split}
\delta R_{ij} =-\frac{1}{2}\ddot{h}_{ij}+\frac{\dot{a}}{2a}&(\dot{h}_{ij}-h_{kk}\delta_{ij})+ \\  + \frac{\dot{a}^2}{a^2}&(-2h_{ij}+ \frac{1}{2}h_{kk}\delta_{ij})+ \\ +\frac{1}{2a^2}(\partial_k\partial_kh_{ij}-&\partial_k\partial_jh_{ki}-\partial_k\partial_i h_{kj}-\partial_j\partial_ih_{kk})
\end{split}
\end{equation}

We consider a perfect fluid whose energy-momentum tensor on a synchronous reference frame is given by $T_{\mu\nu}=\text{diag}(\rho,-P,-P,-P)$ and we can perturb the physical quantities:
\begin{equation}
\rho=\bar{\rho}+\delta\rho \quad P=\bar{P}+\delta P \quad u_{\mu}=\bar{u}_\mu +\delta u_\mu
\end{equation}
Recalling that the normalization of the four-velocity has to be maintained, it is found that $\delta u^0=0$ for a synchronous reference frame where  $u_{\mu}=(1,\vec{0})$ is always a solution of the geodesic equation and the components of the linearized energy-momentum tensor will be:
\begin{equation}
\delta T_{ij}=a^2\delta P \delta_{ij} -\bar{P}h_{ij}
\end{equation}
\begin{equation}
\delta T_{0i}=(\bar{\rho}+\bar{P})\delta u_i
\end{equation}
\begin{equation}
\delta T_{00}= \delta \rho
\end{equation}
so that:
\begin{equation}
\delta S_{ij}= \frac{1}{2}(a^2(\delta \rho-\delta P)\delta_{ij}-h_{ij}(\bar{\rho}-\bar{P}))
\end{equation}
\begin{equation}
\delta S_{0i}=(\bar{\rho}-\bar{P})\delta u_{i}
\end{equation}
\begin{equation}
\delta S_{00}=\frac{1}{2}(\delta \rho +3\delta P)
\end{equation}
We are now ready to write the first order Einstein equations which read:
\begin{equation}
(00): \quad \ddot{h}_{ii}-\frac{2\dot{a}}{a}\dot{h}_{ii}+2\Big(\frac{\dot{a}^2}{a^2}-\frac{\ddot{a}}{a}\Big)h_{ii}=\chi a^2(\delta \rho +3\delta P)
\end{equation}
\begin{equation}
(0i): \quad \frac{1}{2}\partial_t\big(\frac{1}{a^2}(\partial_ih_{jj}-\partial_j h_{ij})\big) =\chi (\bar{\rho}-\bar{P})\delta u_{i}
\end{equation}
\begin{equation}
\begin{split}
(ij): \quad  & \ddot{h}_{ij}-\frac{\dot{a}}{a}(\dot{h}_{ij}-h_{kk}\delta_{ij})+2\frac{\dot{a}^2}{a^2}(2h_{ij}-h_{kk}\delta_{ij}) +\\ & -\frac{1}{a^2}(\partial_k\partial_kh_{ij}+\partial_i\partial_jh_{kk}-\partial_i\partial_kh_{kj}-\partial_k\partial_jh_{ik})=\\ & \chi(-a^2(\delta \rho- \delta P)\delta_{ij}+h_{ij}(\bar{\rho}-\bar{P}))
\end{split}
\end{equation}

The above equations take into account three types of modes: scalar, vector and tensor modes. \\
With another infinitesimal coordinate transformation, we impose the TT gauge that is obtained by the conditions $h_{kk}=0$ and $h_{ij,j}=0$, where repeated indices have to be summed over. In this gauge only the tensor modes, in which we are interested, appear and we obtain the following equation of propagation for the gravitational waves:
\begin{equation}
\ddot{h}_{ij}-\frac{\dot{a}}{a}\dot{h}_{ij}+4\frac{\dot{a}^2}{a^2}h_{ij}-\frac{1}{a^2}\partial_k\partial_kh_{ij}=\chi h_{ij}(\bar{\rho}-\bar{P})
\end{equation}

Using the acceleration \eqref{F2std(a)} and Friedmann \eqref{F1std(a)} equations for the flat FLRW, the equations above can be rewritten as follows \cite{book:weinberg}:
 \begin{equation}
 \label{equazione non riscalata}
 \ddot{h}_{ij}-\frac{\dot{a}}{a}\dot{h}_{ij}-2\frac{\ddot{a}}{a}h_{ij}-\frac{1}{a^2}\partial_k\partial_kh_{ij}=0
 \end{equation}
 which can be again written for a monochromatic wave with wave number $k$:
 \begin{equation}
 \ddot{h}_{ij}-\frac{\dot{a}}{a}\dot{h}_{ij}-2\frac{\ddot{a}}{a}h_{ij}+\frac{k^2}{a^2}h_{ij}=0
\end{equation}

Since the physical wave has to satisfy the condition: 
\begin{equation}
\Big|  \frac{h_{\mu\nu}}{\bar{g}_{\mu\nu}} \Big| \ll 1
\end{equation}
i.e. the condition of smallness compared to the background which becomes:
 \begin{equation}
\Big| \frac{h_{ij}}{a^2} \Big| \ll 1
\end{equation}
for a synchronous reference frame, it is found that the physical wave is the rescaled one, $\tilde{h}_{ij}=\frac{h_{ij}}{a^2}$, for which the propagation equation reads as:
\begin{equation}
\label{eq:equazioneriscalata}
\ddot{\tilde{h}}_{ij}+3\frac{\dot{a}}{a}\dot{\tilde{h}}_{ij}+\frac{k^2}{a^2}\tilde{h}_{ij}=0
\end{equation}

\begin{figure}[h]
\includegraphics[scale=0.70]{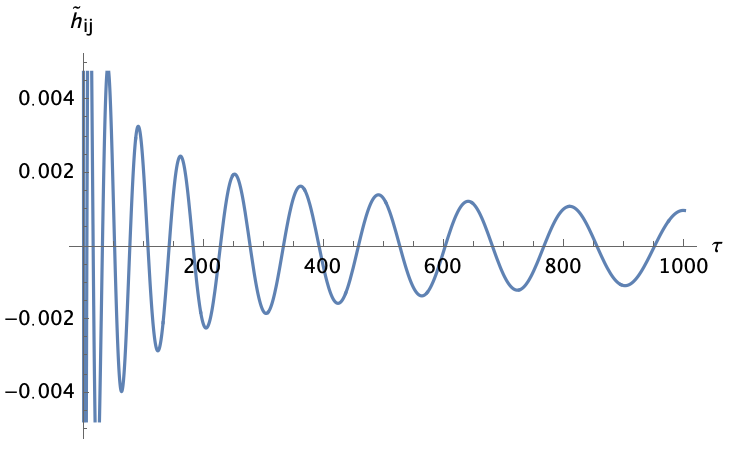}
\caption{Solution for the rescaled perturbation  $\tilde{h}_{ij}$ with wave number $k t_P=1$, where $t_P$ is the Planck time, as function of the evolution parameter $\tau=\frac{t}{t_P}$.}
\label{fig:ondariscalata}
\end{figure}

The solution, displayed for the radiation dominated Universe ($a(t) \propto \sqrt t$) in fig. \ref{fig:ondariscalata}, appears as a damped oscillator with a divergence toward the singularity. This shows that, just as any other physical quantity, even the  amplitude of gravitational waves diverges toward the Big-Bang, in particular as the amplitude grows larger than one, perturbation theory can no longer be valid and the model is non-predictive. 

\begin{figure}[h]
\includegraphics[scale=0.70]{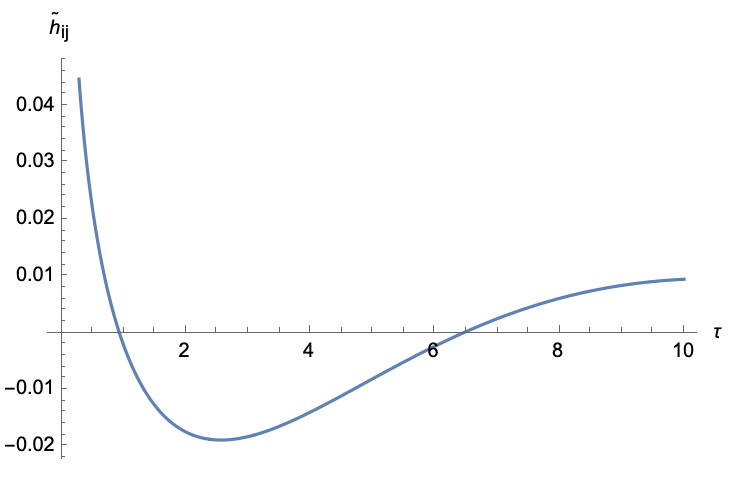}
\caption{Solution for the rescaled perturbation  $\tilde{h}_{ij}$ with wave number $k t_P=1$, where $t_P$ is the Planck time, as function of the evolution parameter $\tau=\frac{t}{t_P}$, the divergence toward the singularity is shown.}
\label{fig:ondariscalata2}
\end{figure}

\subsection{Gravitational waves on a semiclassical Polymer flat FLRW background}
As shown previously in section \ref{sec:exact}, the modification of the FLRW metric by the semiclassical Polymer theory causes the singularity to be removed so that a Bounce, i.e. a minimum for the scale factor, appears. It is then expected that, just as any other physical quantity, also the amplitude of the gravitational waves in such a background will be regularized. To see this, it is first necessary to write the equation of propagation in terms of the volume $V$ for the monochromatic wave:
\begin{equation}
\ddot{h}_{ij}-\frac{1}{3}\frac{\dot{V}}{V}\dot{h}_{ij}+\frac{4}{9}\frac{\dot{V}^2}{V^2}h_{ij}+\frac{k^2}{V^\frac{2}{3}}h_{ij}=\chi h_{ij}(\bar{\rho}-\bar{P})
\end{equation} 
and then in terms of the non dimensional quantities $\tau=\frac{t}{t_P}$, $q=kt_P$, $Q=\frac{\bar{\rho}}{\rho_P}=\frac{1}{V^\gamma}$, where $\rho_P=\frac{8\pi}{t_P^2\chi}$ and $\gamma$ is the polytropic constant:
 \begin{equation}
 \begin{split}
&\frac{d^2{h}_{ij}}{d\tau^2}-\frac{1}{3}\frac{1}{V}\frac{dV}{d\tau}\frac{d{h}_{ij}}{d\tau} +\\ +&\frac{4}{9}\frac{1}{V^2}\Bigg(\frac{dV}{d\tau}\Bigg)^2h_{ij}+\frac{q^2}{V^\frac{2}{3}}h_{ij} -8\pi Q(2-\gamma) h_{ij}=0
\end{split}
\end{equation}

Next we write the equation for the radiation dominated universe, but this time using the Polymer modified Friedmann equation \eqref{polyF1}:
 \begin{equation}
 \begin{split}
& \frac{d^2{h}_{ij}}{d\tau^2}\mp\frac{1}{3} \sqrt{24\pi Q\Big(1-\frac{Q}{Q_\mu}\Big)}
\frac{d{h}_{ij}}{d\tau} \\ &+\frac{32}{3}\pi Q\Big(1-\frac{Q}{Q_\mu}\Big)
h_{ij}+\frac{q^2}{V^\frac{2}{3}}h_{ij} -\frac{16\pi}{3}\frac{1}{V^\frac{4}{3}}  h_{ij}=0
\label{eq:onde}
\end{split}
\end{equation}

Where the equation with the minus sign concerns the expanding Universe while the equation with the plus sign concerns the collapsing Universe. With appropriate initial conditions, eq. \eqref{eq:onde} can be solved numerically and the two branches can be joined. The results that will be described are those of the rescaled perturbation, $\tilde{h}_{ij}=\frac{h_{ij}}{V^{2/3}}$, for it represents the physical wave.

\begin{figure}[h]
\includegraphics[scale=0.60]{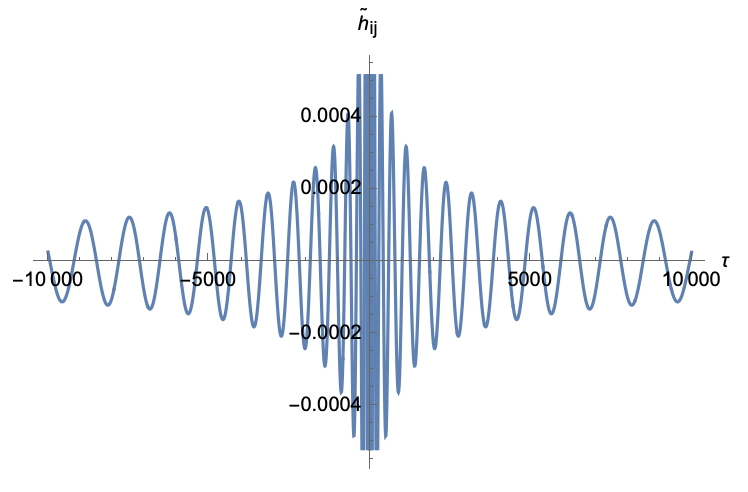}
\caption{Solution for the rescaled perturbation  $\tilde{h}_{ij}$ with wave number $k t_P=1$ and $Q_\mu=1$, where $t_P$ is the Planck time, as function of the evolution parameter $\tau=\frac{t}{t_P}$.}
\label{fig:onda1}
\end{figure}

\begin{figure}[h]
\includegraphics[scale=0.60]{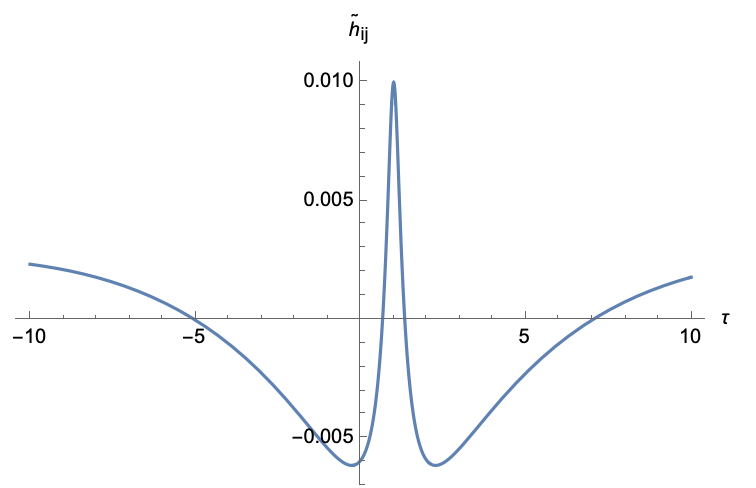}
\caption{Solution for the rescaled perturbation  $\tilde{h}_{ij}$ with wave number $k t_P=1$ and $Q_\mu=1$, where $t_P$ is the Planck time, as function of the evolution parameter $\tau=\frac{t}{t_P}$, the elimination of the singularity at the Bounce is clear in this graph.}
\label{fig:onda2}
\end{figure}

It is seen in Fig. \ref{fig:onda1} that both branches are damped oscillators that grow toward the time of the Bounce where, as expected, the amplitude takes a finite value as shown in Fig. \ref{fig:onda2}. Clearly, this could mean that gravitational waves can propagate from the collapsing Universe to the expanding one through the Bounce and measures of gravitational waves that come from events happened before the Bounce could be made possible.  This theoretically implies that we could get information about the Universe before the time of the Bounce. This is not possible by means of electromagnetic measures.

It also needs to be specified that the described model holds only when perturbation theory holds, i.e. when the amplitude of the gravitational wave in question remains smaller than one; if this condition is not satisfied, Einstein equations must be solved exactly. Nonetheless, the model remains valid for all those perturbations that are born close enough to the Bounce, that they wouldn't have time to grow in amplitude and exit the range of validity of perturbation theory. 


We can also study how varying the parameters $q$ and $Q_\mu$ changes the appearance of the solution.

Keeping in mind that eq. \eqref{eq:onde} looks like a damped oscillator with non constant coefficients, it is clear that the period of such oscillations will depend on the inverse of the wave number, in a way that the greater is $q$ the denser the oscillation will be, as shown in Fig. \ref{fig:onda3}.

On the other hand, changing the parameter $Q_\mu$ corresponds to changing the Polymer scale $\mu_0$ and in the limit $Q_\mu \rightarrow \infty$ the classical limit is recovered. For this reason, we expect that by increasing the value of $Q_\mu$ the solution reaches greater values around the Bounce so that it can recover the divergence in the true classical limit.

Furthermore, the above results are compatible with the analysis carried out in \cite{art:lumin} where the gravitational wave amplitude today as a function of the Bounce temperature is plotted. Here it is found that such amplitude reaches an asymptotic value as long as a high enough temperature is assumed.  This asymptotic value must be very close to the classical one since, as we go far from the Bounce, polymer effects are negligible, and, for this reason, the value of the minimal volume, related to the Bounce temperature, will not influence the amplitude of the gravitational wave detected today.

\begin{figure}[h]
\includegraphics[scale=0.60]{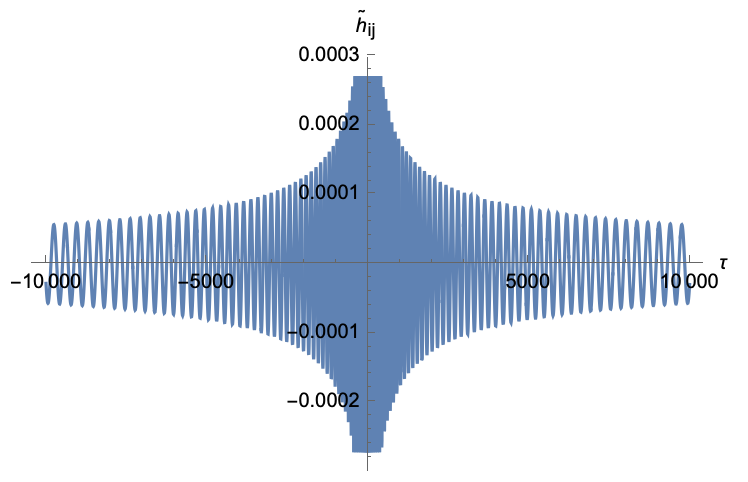}
\caption{Solution for the rescaled perturbation  $\tilde{h}_{ij}$ with wave number $k t_P=\sqrt{17}$ and $Q_\mu=1$, where $t_P$ is the Planck time, as function of the evolution parameter $\tau=\frac{t}{t_P}$. A greater wave number causes the oscillation to get denser.}
\label{fig:onda3}
\end{figure}
\begin{figure}[h]
\includegraphics[scale=0.60]{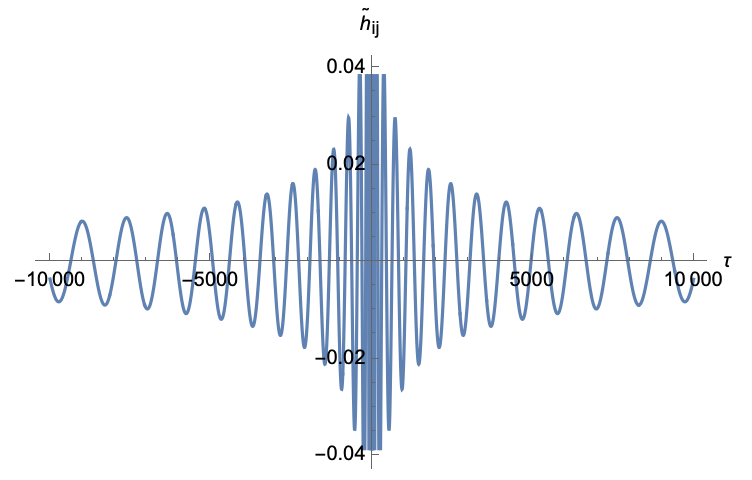}
\caption{Solution for the rescaled perturbation  $\tilde{h}_{ij}$ with wave number $k t_P=1$, where $t_P$ is the Planck time, as function of the evolution parameter $\tau=\frac{t}{t_P}$. Here, it has been chosen the parameter $Q_\mu=10^5$ and it can be seen that the solution grows larger near the Bounce compared to the solution with smaller $Q_\mu$.}
\label{fig:onda3}
\end{figure}
\subsection{Spectral dependence and Gaussian wave packet}
\begin{figure}[h]
\includegraphics[scale=0.60]{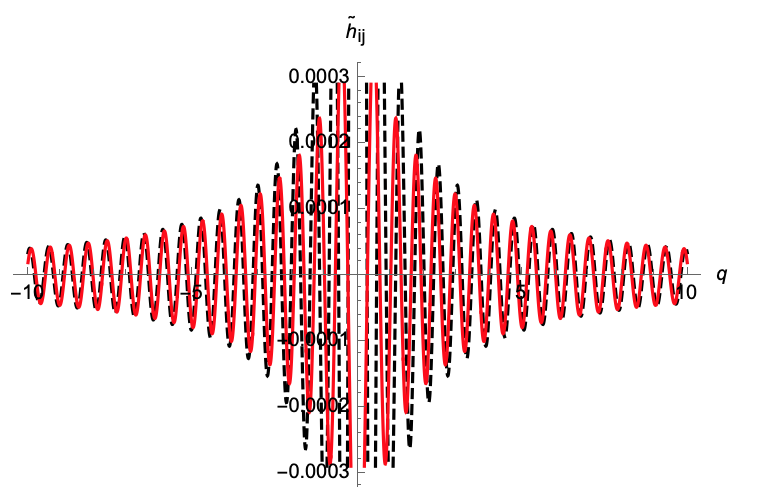}
\caption{Classical (dashed) and polymer (continuous) amplitude, $\tilde{h}_{ij}$, as a function of $q=kt_P$ at the fixed time $\tau=2$, with $Q_\mu=1$.}
\label{fig:onda5}
\end{figure}

\begin{figure}[h]
\includegraphics[scale=0.60]{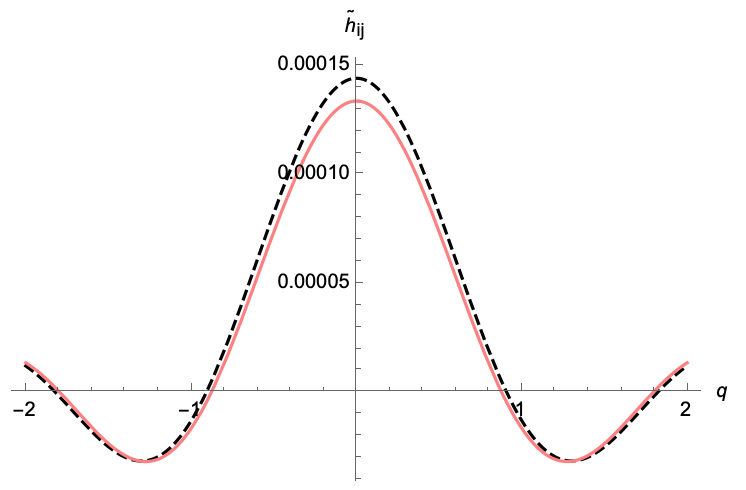}
\caption{Classical (dashed) and polymer (continuous) amplitude, $\tilde{h}_{ij}$, as a function of $q=kt_P$ at the fixed time $\tau=100$, with $Q_\mu=1$.}
\label{fig:onda6}
\end{figure}

We can now study the dependence on the non dimensional wave number,  $q=kt_P$, of the amplitude at a fixed time, i.e. the spectral dependence of the amplitude. The latter corresponds, once again, to a damped oscillator, both for negative and positive wave numbers, that grows as $q\rightarrow 0$. This is the case also for the classical solution with which we can compare the polymer one. As we go close to the Bounce, like in Fig. \ref{fig:onda5}, the classical and polymer cases differ in amplitude around small $q$ and become very similar for large $q$. By remembering that the choice $Q_\mu=1$ implies the Polymer scale to be $\sqrt[3]{\mu_0}=O(t_P)$, the above results suggest that polymer effects are visible mostly for those wave lengths that are of the order or greater than the polymer scale itself. Furthermore, by looking at Fig. \ref{fig:onda6}, it is clear that at times far away from the Bounce the polymer and classical waves tend to be similar even at small $q$ and we find a classical limit for large times, as we expected. 

\begin{figure}[t]
\includegraphics[scale=0.60]{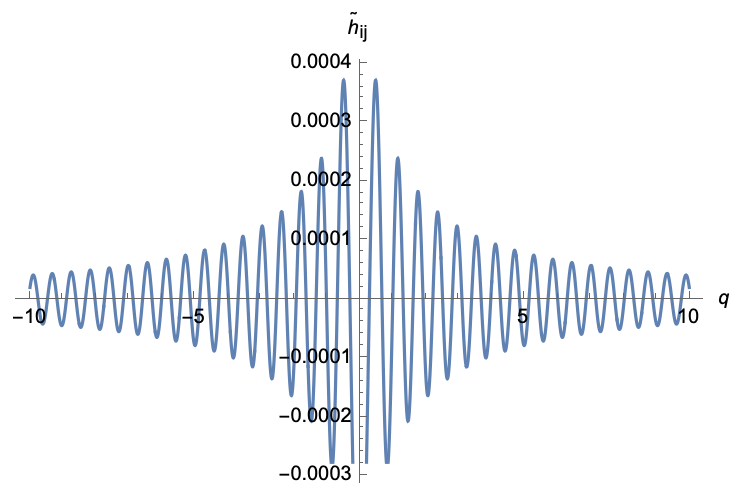}
\caption{$\tilde{h}_{ij}$ as a function of $q=kt_P$ at the fixed time $\tau=2$, with $Q_\mu=1$.}
\label{fig:onda4}
\end{figure}

It is then useful to look at the time evolution of a Gaussian wave packet during the history of the Universe. For this reason, we can consider a Gaussian wave packet on the domain of wave numbers and, by doing the one dimensional Fourier Transform, it is possible to find the evolution of the packet in the domain of position. Such a Fourier Transform is given by:
\begin{equation}
\bar{h}_{ij}(x,\tau)=\int \frac{dq}{2\pi}\tilde{h}_{ij}(q,\tau)g(q)e^{-iqx}
\end{equation} 
where $\tilde{h}_{ij}(q,\tau)$ is the solution of the wave equation and $g(q)$ is the Normal distribution:
\[
g(q)=\frac{1}{\sigma \sqrt{2\pi}}e^{\frac{-(q-\tilde q)^2}{2\sigma^2}}
\]
with standard deviation $\sigma$ and mean value $\tilde q$. The results are computed numerically and displayed below in Fig. \ref{fig:poly} for three different times: $\tau=2,1000,9000$. Such results show that a Gaussian wave packet in the domain of the comoving wave numbers corresponds to a Gaussian packet in the domain of position and it doesn't spread throughout the history of the Universe although it does lower in amplitude. Furthermore, since the shape of $\tilde{h}_{ij}(q,\tau)$ is symmetric for the collapsing and expanding Universes, also the appearance of the wave packet will be symmetric.

\begin{figure}[h]
\centering
\subfloat[][\emph{$\tau=2$}.]
{\includegraphics[scale=0.6]{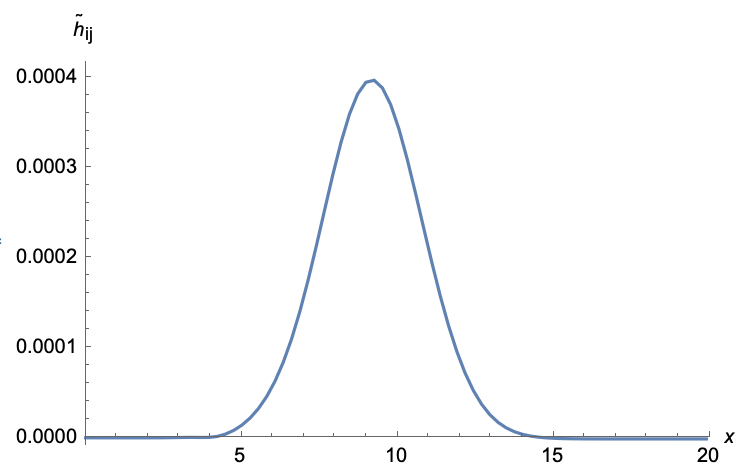}} \\
\subfloat[][\emph{$\tau=1000$}.]
{\includegraphics[scale=0.6]{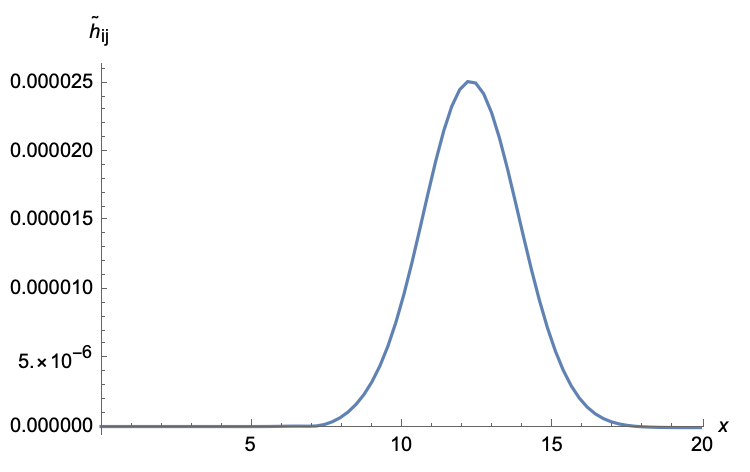}} \\
\subfloat[][\emph{$\tau=9000$}.]
{\includegraphics[scale=0.6]{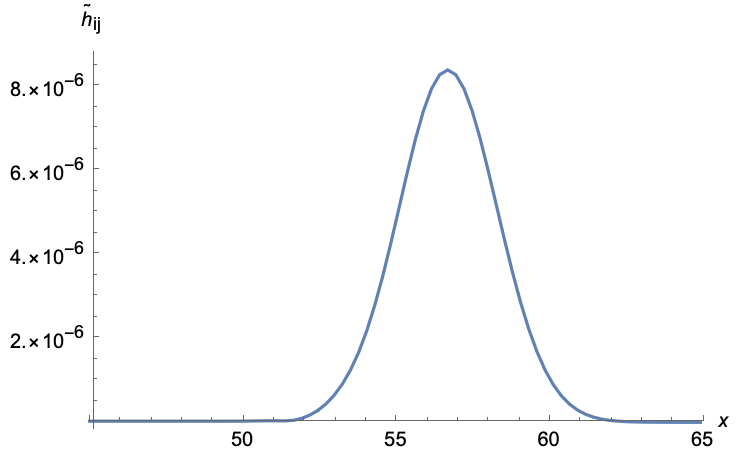}}
\caption{Time evolution of a Gaussian wave packet with parameters $\sigma=0.5$ e $\mu=2$, in a classical FLRW background.}
\label{fig:classico}
\end{figure}

In Fig. \ref{fig:classico} a classical wave packet is shown and it can be noted that the evolution looks very similar to the polymer case, but, as expected,  the amplitudes will evolve differently.

\begin{figure}[h]
\centering
\subfloat[][\emph{$\tau=2$}.]
{\includegraphics[scale=0.6]{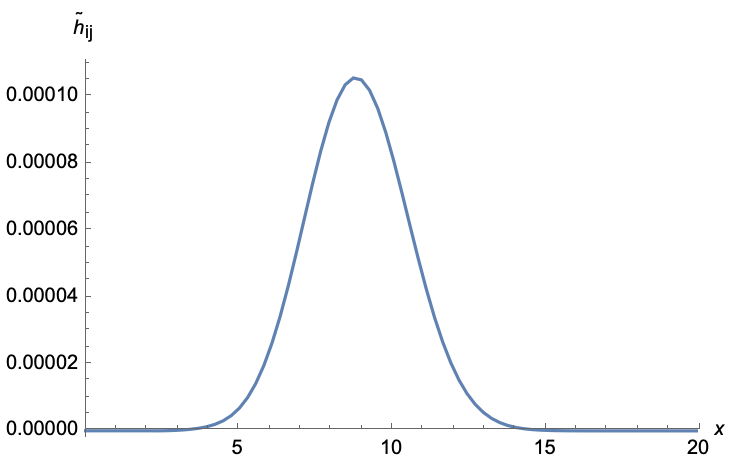}} \\
\subfloat[][\emph{$\tau=1000$}.]
{\includegraphics[scale=0.6]{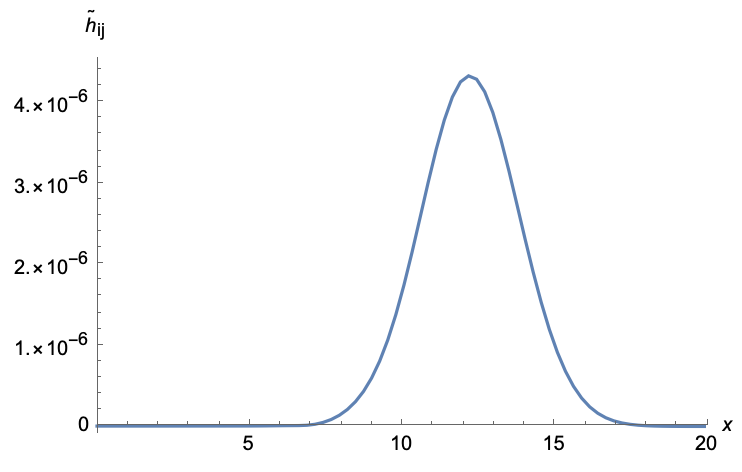}} \\
\subfloat[][\emph{$\tau=9000$}.]
{\includegraphics[scale=0.6]{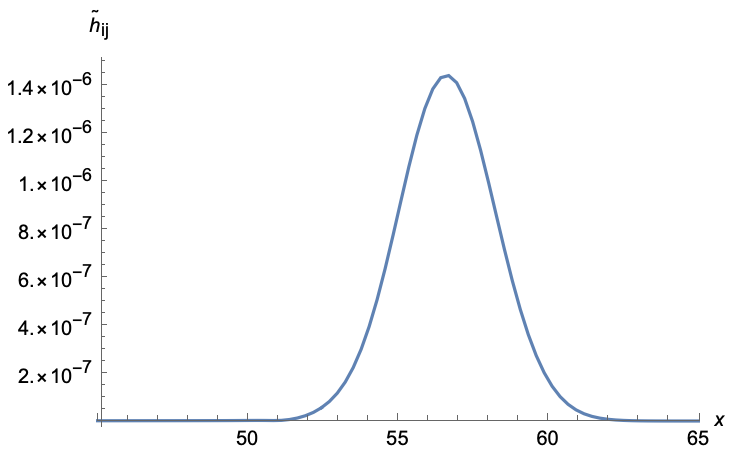}}
\caption{Time evolution of a Gaussian wave packet with parameters $\sigma=0.5$ e $\mu=2$, in a polymer-modified FLRW background.}
\label{fig:poly}
\end{figure}

\section{Concluding remarks} \label{conclusions}
In this paper, we have developed a detailed analysis of phenomenological effects concerning the flat isotropic Universe as described in a semiclassical polymer representation.

We start by studying the perturbative case of a polymer generalized Hamiltonian dynamics, as restated via modified Poisson brackets. This paradigm allows a parallelism with the so-called Generalized Uncertainty Principle formulation, associated, like the polymer procedure, to a minimal cut-off scale, but being thought of as the low-energy limit of String or Brane Theory. We show that the parallelism is directly reflected into a different structure of the Friedmann equation: the different signs of the two approaches correspond to two different signs in the term modifying the matter source. While the perturbative polymer is associated to a non-singular Cosmology (exactly as the full polymer scenario and the semiclassical Loop Quantum Cosmology), the Generalized Uncertainty Principle still predicts a Big Bang Cosmology (for a discussion on quantum level see \cite{art:FLRWGUP}), just like the Brane Cosmology Friedmann equation. Clearly, the modified Poisson brackets approach gives a matter source that is obtained from the exact one (semiclassical Loop Quantum Cosmology and Brane Cosmology), when the ratio between the Universe energy density and its critical value is expanded up to first order.

This analysis suggests that, as far as we remain well below the cut-off energy density, we can use the modified Poisson brackets approach, leaving free choice on the two signs, and then qualitatively describe String and Loop Cosmology, respectively.

An interesting point that could be addressed in a future analysis is how the background dynamics obtained with the perturbative polymer approach differs from the standard LQC results reported, for example, in \cite{art:ashtekarwmap, art:bianchi1inflation} where the cosmological Klein-Gordon equation is solved for a massless scalar field in order to derive the duration of the inflationary phase compatible with the observed data. However, it is worth stressing that, as suggested by the smallness of the tensor to scalar ratio in the primordial perturbations \cite{art:planck2018}, the inflationary phase has to take place in a purely classical region of the cosmological dynamics. Thus the cut-off physics effects on the background evolution are expected to be essentially negligible when the de Sitter phase starts at temperature values three of four orders smaller than the Planck scale. Of more physical impact could be the study of the deformation of the scalar perturbation spectrum induced by the implementation of Polymer Quantum Mechanics in the scalar field fluctuations, see for instance the analysis in \cite{art:polyperturb}

Then, we face the subtle question about the possibility to regularize the vacuum energy density of a free massless scalar field by a polymer regularization of its second quantization. 

Actually, we arrive to a finite value of the vacuum energy density of the scalar field in second quantization, although a generalization of the creation and annihilation operators is not possible. 

The main merit of this analysis consists in the constant character of the vacuum energy density which lives on a flat isotropic Universe, whose dynamics is seen in the semiclassical polymer representation. In other words, we demonstrate the emergence of a Cosmological Constant from the regularized vacuum energy of the scalar field. Clearly, if the discretization parameter in the adopted configurational variable, i.e. the Universe volume, is taken of the order of the Planck scale, the obtained Cosmological Constant is still of the cut-off order and it is unable to account for the present value of the Cosmological Constant, presumably accelerating the present day Universe. Yet, a discussion about the fine-tuning of the model parameters which provides the right value of the present Universe acceleration is also discussed.

However, we raise the question about the non-stationary character of such a vacuum energy density, as a consequence of the fact that the corresponding Hamiltonian function is time dependent and therefore its eigenstates are not physical states. Actually, the observed vacuum energy of the Universe is determined by the projection of a generic physical state on the vacuum state here discussed. This problem could be numerically addressed only once all the excited states have been regularized, according to the polymer procedure, facing a non-trivial mathematical problem. Hence, a numerical analysis with different initial conditions can be performed to clarify the real phenomenology of the mean value associated to this Cosmological Constant.

Finally, we investigate the question concerning the behaviour of gravitational waves living on the semiclassical polymer flat isotropic Universe. We analyze the deformation of the wave amplitude and spectrum and we demonstrate that the presence of a Bounce prevents the divergence of the wave amplitude, which is typical of the Big Bang Cosmology. 

In principle, we could observe gravitational waves emitted before the Bounce during the collapsing process of the Universe. In fact, sufficiently small space time ripples, produced during the pre-Big Bounce phase (for instance by galaxy crunches), could reach the turning point of minimal volume, still in the linear regime, and then they could, in principle (i.e. if not thermalized), reach our Earth and current and future detectors. This possibility opens an intriguing perspective on the chance to search for information of the pre-Big Bounce Universe, currently propagating in our expanding branch. 

The present formulation of the cosmological dynamics in a polymer semiclassical representation, closely resembling the Big-Bounce features of LQC, calls for a full quantum implementation of the considered scheme in order to analyze the cosmological perturbation spectrum. The aim could be to understand if Polymer Quantum Mechanics is able to provide results similar to those predicted in \cite{art:bojoprl}, where features of the cosmological perturbation spectrum are investigated in the LQG framework. Since our polymer analysis is based on a metric approach the methodology that could be addressed is analogous to the one proposed in \cite{art:kief2016}, for the standard WKB formulation to the Wheeler-de Witt approach. However two considerations are in order. Firstly, the methodology proposed in \cite{art:kief2016}, based on an expansion in the Planck mass of the theory, predicts non-unitary Quantum Gravity corrections to Quantum Field Theory; this question  must be primarily addressed before implementing this procedure in the polymer sector (see for instance \cite{art:kief2018}). Secondly, the semiclassical corrections to a De Sitter phase of inflation, as described by the polymer formulation, are expected to be very small because inflation takes place rather far from the planckian epoch and the cut-off physics effects are essentially vanishing. Nonetheless, once fixed a predictive unitary formulation of quantum perturbation dynamics, it would be relevant to perform a study of Quantum Gravity corrections to the spectrum of fluctuations that led to the late Universe structure formation.

Altogether, the results discussed in this manuscript show how the presence of the Bounce, due to a cut-off physics, can alter our investigation and interpretation of the present Universe. The presence of a non-singular turning point of minimal volume in the past of our Universe does not correspond simply to a cut-off on a diverging energy density, making the theory of the Big Bounce physical, but it also gives a completely new point of view on the present cosmological phenomenology.

%

\end{document}